\documentclass[12pt]{article}
\usepackage[centertags]{amsmath}
\usepackage{fancyhdr}
\usepackage[bottom]{footmisc}

\usepackage{amsfonts}
\usepackage{amssymb}
\usepackage{amsthm}
\usepackage{newlfont}
\usepackage[english]{babel}
\usepackage{graphicx}
\usepackage{array}
\usepackage{enumitem}

\usepackage{titlesec}

\titleformat{\section}
{\normalfont\large\bfseries}
{\thesection. }
{0pt}
{}

\numberwithin{equation}{section}

\begin{document}

\def\baselinestretch{1.7}

\title{The  $SU(r)_2$  string functions as $q$-diagrams} \vspace{-0.1in}
\author{Arel Genish \& Doron Gepner \vspace{0.1in}\\
\small\textit{{Department of Particle Physics, Weizmann Institute, Rehovot, Israel}}   \vspace{-0.1in} \\
\small\textit{{E-mail: Arel.genish@weizmann.ac.il, Doron.gepner@weizmann.ac.il}}
\date{\small{\today}}}

\maketitle

{\center ABSTRACT \\} 
 \vspace{0.1in} 
A GRR expression for the characters of $A$-type parafermions has been a long standing puzzle dating back to conjectures made regarding some of the characters in the 80's. Not long ago we have put forward such GRR type identities describing any of the level two $ADE$-type generalized parafermions characters at any rank. These characters are the string functions of simply laced Lie algebras at level two as such they are also of mathematical interest. In our last joint paper we presented the complete derivation for the $D$-type generalized parafermions characters identities. Here we generalize our previous discussion and prove the GRR type expressions for the characters of $A$-type generalized parafermions. To prove the $A$-type GRR conjecture we further our study of $q$-diagrams, introduced in our last joint paper, and examine the diagrammatic interpretations of known identities among them Slater identities for the characters of the first minimal model and the Bailey lemma.

\section{\large{Introduction}}
Two dimensional theories comprised of a matter content which includes generalized parafermions have been the subject of many a papers along the years. A prominent example that has attracted a lot of interest since the surfacing of AGT correspondence \cite{AGT} is the $N$'th affine para-Toda theory\cite{PT1,PT2}. The arrival at this models has been the result of some progress. First it was realised in \cite{f4,f1,f2}  that CFTs with affine and $W_k$-symmetry are related to the instanton counting for the $SU(r)$ gauge group. This led the authors of \cite{f} to extend the correspondence between instanton partition functions and conformal blocks of two-dimensional CFT's to the case of $N = 1$ SUSY. In particular, it was found that the instanton partition function of the $SU(2)$ Yang Mills theory evaluated on the $\mathbb{Z}_2$ symmetric instanton moduli space is related to super Liouville conformal blocks in the Whittaker limit. Interestingly, the symmetry of this model is the affine $SU(2)$ at level two and the super-Virasoro symmetries. In search of further generalizations of the $4$-d instanton partition function  $2$-d CFT correspondence web a further idea was proposed in \cite{M5}.
Following $M$-theory interpretation of two $M5$-branes on $\mathbb{R}_4/\mathbb{Z}_2$ it was suggested that $k$ $M5$-branes on $\mathbb{R}_4/\mathbb{Z}_N$ realize a $2d$ theory with a free boson, the affine $SU(N)_k$ , and the $N$-th para-$W_k$ symmetry. Naturally, this includes the standard $W_k$ symmetry for $N= 1$ and super-Viraroro symmetry for $k = N = 2$ just mentioned. Finally, the $N$-th para-$W_k$ algebra is the symmetry of the $N$-th para-Toda model of type $SU(N)$, which has the action \cite{S}
\begin{eqnarray}\label{1.1}
S(\frac{SU(r+1)_k}{U(1)^r})+\int d^2x [\partial_{u}\Phi\partial_{u}\Phi + \sum^{r}_{i=0}\Psi^{0}_{\alpha_i}\bar{\Psi}^{0}_{\alpha_i}exp(\frac{b}{\sqrt{k}}\alpha_i \cdot \Phi)]
\end{eqnarray}
where $\Phi$ is a vector of $r$ boson fields, $\alpha_i$ are the simple roots of the affine $SU(r+1)_k$, $b$ is related to the background charge and $S(\frac{SU(r+1)_k}{U(1)^r})$ stands for the formal action of the generalized parafermions $\Psi^{\Lambda}_{\lambda}$. As is implied above, generalised $G(r)_k$ type parafermions\footnote{These were separately developed in mathematic as $Z$ algebras \cite{lep}} $\Psi^{\Lambda}_{\lambda}$ are more generally defined  as describing the excitations associated to the 
\begin{eqnarray}\label{1.2}
H(G(r)_2)=\frac{G(r)_k}{U(1)^r}
\end{eqnarray}
coset CFT \cite{gep1} where, for our purposes, $G(r)_k=A(r)_k,D(r)_k,E(r)_k$ is any of the simply laced affine Lie algebras of rank $r$ and level $k$. \\
A daunting problem in the study of such theories, which include generalized parafermions in the matter content, is describing the partition function of these theories and in particular the characters associated with the primary generalized parafermions. Indeed, until recently, the characters corresponding to the generalized parafermion primaries
were actually unknown. In a series of articles initiated by one of the authors and A.Belavin these have gradually been uncovered. First, the characters of $SU(N)_2$ generalized parafermions were found in \cite{belgep} via the ladder coset construction. Interestingly, it was shown that the $A$ type parefermions theories of level two and any rank can be realised by a product theory of minimal models with particular combinations of the representing fields taken to insure modular invariance is preserved. This was followed by \cite{gengep} where this program was generalized to all simply laced affine Lie algebras and the characters of ADE generalized parafermions at level two and any rank were also found. More specifically, the authors of \cite{belgep} considered the coset
 \begin{eqnarray}\label{1.3}
\mathcal{A}(k,r)=\mathcal{H} \times SU(r)_k \times \frac{ SU(k)_r \times SU(k)_n}{SU(k)_{r+n}}
\end{eqnarray}
 corresponding to the construction described above for the $k$ $M5$-branes on $\mathbb{R}_4/\mathbb{Z}_N$ instanton partition function where $n$ is given in terms of the Nekrasov parameters $\epsilon_{1,2}$ \cite{nek}. For two $M5$-branes on $\mathbb{R}_4/\mathbb{Z}_N$ it was observed that, up to $U(1)$ factors which enter trivially in the characters, the coset theory $A(k,r)$ is described in a more illuminating fashion via the use of level-rank duality and the ladder coset construction 
 \begin{eqnarray}\label{1.4}
\mathcal{A}(2,r)^{m_{r+1}}_{k_1,...,k_r,k_{r+1}}=\sum_{m_2,...,m_r \atop m_i+m_{i+1}= k_{i+1} \mod 2} \prod_{i=2}^{r} \frac{SU(2)^{m_{i-1}}_{i-1} \times SU(2)^{k_{i}}_1}{SU(2)^{m_{i}}_{i}} \times \frac{SU(2)^{m_{r}}_{r} \times SU(2)^{k_{r+1}}_n}{SU(2)^{m_{r+1}}_{n+r}}
\end{eqnarray}
where we denote by $SU(2)_{f}^{s}$ the affine theory of level $f$ and the representation of twice isospin s, $ 0 \leq s \leq f$. The indices $k_i=0,1$ for $i=1,...,r+1$ and $m_i=0,...,i$, the summation is taken over $m_i$ for $i=2,...,r$ and we find it convenient to define $m_1=k_1$.
for $n=1$ one can immediately identify this model as a product theory of the first $r$ minimal models accordingly the characters of the level two $A$ type generalized parafermions were found to be given by a suitable sum over products of the minimal models characters
 \begin{eqnarray}\label{1.5}
c^{l}_{k_1,...,k_r,k_{r+1}}=\sum_{m_2,...,m_r \atop m_i+m_{i+1}= k_{i+1} \mod 2} \prod_{i=1}^{r} M(m_{i}+1,m_{i+1}+1)_i
\end{eqnarray}
Here the characters of the $i$'th minimal model are denoted by $M_i$ these are well known
 \begin{eqnarray}\label{1.6}
M(n,m)_i=\frac{1}{\eta(q)}\sum_{s=0,1}(-1)^s\Theta_{\lambda_s(n,m),(i+2)(i+3)}(q)
\end{eqnarray}
where,
 \begin{eqnarray}\label{1.7}
\lambda_s(n,m)=n(i+3)-m(i+2)(1-2s)
\end{eqnarray}
and the theta functions at level $h$
 \begin{eqnarray}\label{1.8}
\Theta_{\lambda,h}(q)=\sum_{l \in \mathbb{Z}}q^{h(l+\frac{\lambda}{2h})^2}.
\end{eqnarray}
Finally, using level-rank duality again for $n=1$ in the ladder coset representation eq. (\ref{4.1}) one finds that this coset is equally described by a the $A$-type generalized parafermions theory. \\
Fascinating as this correspondence between the $A$ type generalized parafermions and the product of minimal theories may be, the characters are of a highly non trivial mathematical structure which makes it particularly hard to use them for further applications. Actually,  this type of problem is known in physics and it's origin can be traced to the hexagon model studied by Baxter \cite{bax}, specifically, the one dimensional configuration sum, where he utilised the famous Ramanujan identity to find the local state probabilities. As we now know, the RSOS models one dimensional sums are identical to the characters of the fixed point CFT in the appropriate regime\cite{mel, gep}. This was later considerably further developed and leads to the conjecture that GRR identities exit for every CFT that appears as a fixed point. With this in mind the authors of \cite{belgep} conjectured and numerically verified GRR identities for the $A$ type generalized parafermion characters. Furthermore, an $ADE$ generalization for level two parafermions soon followed and also verified numerically in \cite{gengep}. Finally although, only $ADE$ level two generalized parafermions characters were given exact analytical expressions the corresponding GRR identities led to a conjecture for the characters of generalized parafermions associated with any Lie algebras at any level and rank level \cite{gep2}. \\
The purpose of this paper is to provide a detailed proof for the GRR identities arising for the $A$ type generalized parafermions characters and is the logical continuation of  
our work in \cite{gengep1} where we have proven the GRR identities arising for the level two $D$ type parafermion characters of rank $r$. This was achieved by describing the GRR identities in terms of $q$-diagrams which were introduced as a general mathematical framework encapsulating all the mentioned identities. These $q$ diagrams made of connected nodes and an assortment of external lines, which can be thought of as a generalization of the Dynkin diagrams, highlight the basic structure of their associated expression and can be shown to posses a symmetry, termed $Q$ symmetry. In particular, Dynkin shape $q$ diagrams encapsulate the associated Lie algebra Cartan matrix while $Q$ symmetry can be realised as the associated Lie algebra Weyl symmetry\cite{gengep1}. In terms of $q$ diagrams the GRR identities are represented by a simple correspondence:
{\center \textbf{Character of $G_2$ type parafermion}  $\Leftrightarrow$  \textbf{G-shaped q-diagram}} \\
For the $D$-type parafermion characters this diagrammatic picture provided us with much needed intuition to prove the correspondence which represent a family of new infinite series of GRR identities as well as providing a relatively simple expression for the characters of $D$-type generalized parafermions. Furthermore, the language of $q$-diagrams revealed a deep connection between various well known identities. For example, the triple Jacobi identity and the Roger-Ramanujan identities were interpreted as the first two identities in an infinite diagrammatic series corresponding to $D$ shaped $q$ diagrams with an assortment of external legs which is naturally seen as the diagrammatic extension of these identities. \\
Indeed, as we shall soon see, following the diagrammatic intuition furnishes a way to also prove the $A$-type parafermions GRR identities
which provides some motivation for the still unproven $E$-type generalized parafermions GRR identities. As this program is similar in spirit to the $D$-type GRR identities proof let us quickly recall the steps and highlight the resemblance. The first step involved simplifying the coset model ladder representation of the $D$ type generalized parafermion characters. The $D$-type generalized parafermion coset theory, in a similar fashion to the one presented here, was shown to be equivalent to a product theory of $r-1$ bosons at various radii,
 \begin{eqnarray}\label{1.9}
R_i=\sqrt{2i(i+1)}.
\end{eqnarray}
Indeed, the characters of the bosonic theory are also expressed in terms of a level $h=R^2/2$ theta function presented above. To simplify the coset ladder representation of $r-1$ free bosons it can be described by an equivalent theory of an $r$ dimensional boson propagating on a lattice via the beta method. Actually, as characters of minimal models are also given by the theta function, albeit a subtraction of two such thetas, the beta method can be applied to the $A$-type ladder coset. This is the subject of section 3 where we show that the product theory of minimal models can be place on a lattice $\tilde{A}$ which, as it turns out, is a simple extension of the $A_r$ root lattice. Indeed, this step is crucial for our analysis and reveals a deeper relation to the $D$ shaped diagrams which in turn  will provide us with the main intuition for proving the $A$-type identities. Next, as in the $D$-type case we will identify the GRR identities needed to prove our correspondence by giving diagrammatic interpretations for known identities. As these diagrammatic interpretations evidently provide a strong tool section 4 is devoted to the diagrammatic interpretation of the Bailly Lemma. As this lemma is basically a mechanism for generating GRR identities the diagrammatic interpretation of such a lemma is of particular interest and potentially leads to an infinite number of new diagrammatic interpretations and extensions of known identities. After deriving the diagrammatic Bailey lemma, we proceed to examine the Bailey pair used by Slater to prove the first $A$-type GRR identity corresponding to the character of the identity in the first minimal model, more prominently known as the Ising model. Extending Slater's derivation is quite tedious and might appear quite ad hoc if it were not for the diagrammatic interpretation of the Bailey lemma which makes it extremely clear conceptually. Finally, the last two sections concentrate on generalizing the results to all characters of the $A$-type generalized parafermions theory\footnote{The reader might recall that the mentioned correspondence, refers strictly to characters associated with fundamental weights. For the case at hand we note that these provide all the characters of the theory due to identifications via the external automorphism of $SU(r)$.}

\section{\large{$q$-diagrams}}

In our last joint paper \cite{gengep1} we have introduced $q$-diagrams as a tailor made tool to prove the GRR identities corresponding to the $SO(2r)_2$ string functions. Furthermore, we have motivated the use of $q$-diagrams to study the level two string functions for any simply laced algebra. One particular nice feature of $q$-diagrams is their associated $Q$-symmetry. In particular, for Dynkin $q$-diagrams corresponding to any Lie algebra it was shown that this symmetry can be realised as the corresponding Weyl symmetry. This already implies that the $SU(r+1)$ $q$-diagrams indeed have the right symmetry structure as to the describe the $SU(r+1)$ level two string functions. In our work regarding the $SO(2r)$ diagrams it was evident that the diagrammatic picture for the string functions gives highly non-trivial intuition as to how can one attack the problem at hand. Let us recall the diagrammatic rules for constructing $q$-diagrams. Using the $G_r$ Dynkin diagram we introduce a set of diagrammatic rules. First, assign for each node at the Dynkin diagram some "momenta" $b_i$ such that $i$ corresponds to the number of the node. In addition, assign a momenta $\Lambda_i$ for each external line connected to the $i$'th node. Next, we prescribe a set of diagrammatic rules, 
\begin{enumerate}[label=\roman{*}., ref=(\roman{*})]
\item \text{for each node} $=\ {q^{b_{i}^2/2}\over (q)_{b_i}}$.
\item \text{for each internal line connecting the $i$'th and $j$'th nodes} $=q^{-b_ib_j/2}$.   
\item \text{for each external line of momenta $\Lambda_{i}$ connected to the $i$'th node} $=q^{-\Lambda_ib_i/2}$.   
\item \text{for each dashed line connecting $b_i$ and $b_j$ } $=q^{b_ib_j/2}$.   
\item \text{sum over all nodes momentas }
$=\sum_{ b_i= 0 }^\infty\frac{1}{2}(1+(-1)^{b_i+Q_i})$.
\end{enumerate}
Where, for now, let us consider $\Lambda=\sum \Lambda_i\omega_i$ any weight with integer Dynkin labels  greater or equal to zero while $Q=\sum Q_i \alpha_i$ is any root vector of $G_r$ and $(q)_b$ is the $q$ Pochhammer symbol. Following these diagrammatic rules one can easily construct the expressions corresponding to any simply laced Lie algebra Dynkin shaped $q$-diagrams. These are simply denoted by $G_r(\Lambda,Q)$, which in the context of $q$-diagrams specifies the shape or internal momenta, length, external momenta and parity restriction of the corresponding $q$-diagram. For our current purpose consider the $q$-diagrams corresponding to the $SU(4)$ algebra with an external momenta corresponding to the second fundamental weight 
\begin{equation}\label{2.1}
\includegraphics[width=0.5\linewidth,keepaspectratio=true]{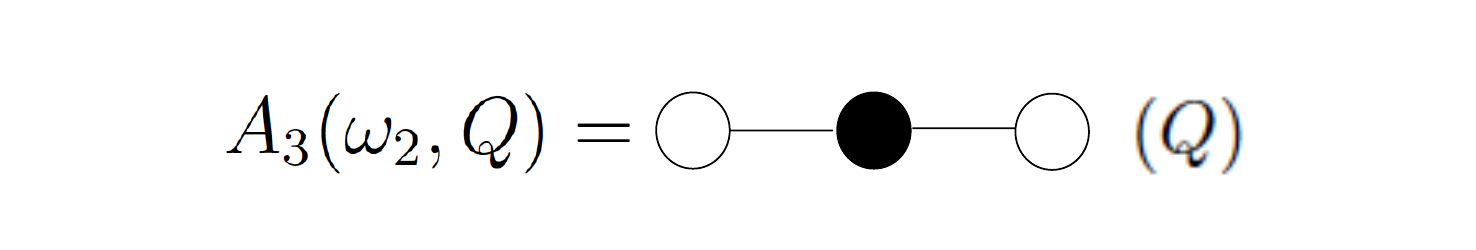}
\end{equation}
where the diagram contains $3$ nodes of which the second node is blacked as a short hand notation for an external momenta corresponding to a fundamental weight $\omega_i$. Following the diagrammatic rules the corresponding fermionic sum is given by
\begin{eqnarray}\label{2.2}
A_3(\omega_i,Q)=\sum_{ b_i=0\atop  b_i= Q_i \mod2 }^\infty 
{q^{ \frac{1}{2}(b_1^2-b_1b_2+b_{2}^2-b_2b_3+b_{3}^2-b_2) }\over (q)_{b_1}(q)_{b_2}(q)_{b_3}},
\end{eqnarray}
here the summation is taken over all  nodes for non negative integers with parity restriction specified by the $Q$ root vector, i.e $b_i \in Q_i+2\mathbb{Z_{\geq 0}}$. \\
In diagrammatic language our conjecture for the level two simply laced string functions boils down to a correspondence between the $SU(r+1)$ $q$-diagrams and the $H(SU(r+1))$ coset theory characters namely,
\begin{equation}\label{2.3}
 \includegraphics[width=0.5\linewidth,keepaspectratio=true]{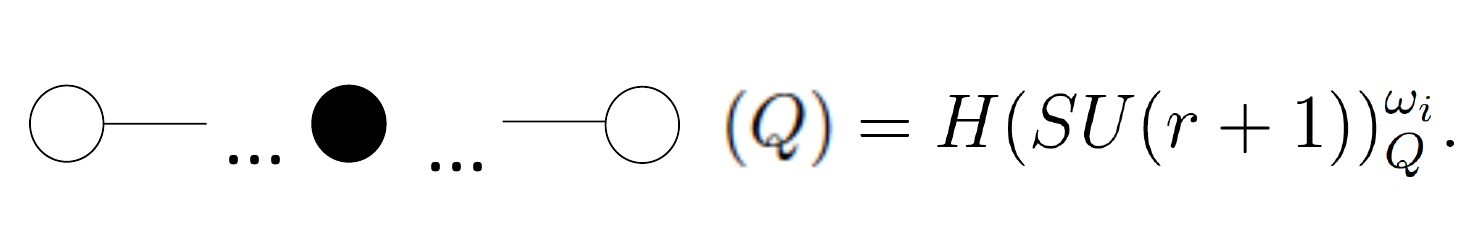}
\end{equation}
where the diagram on the left hand side contains $r$ nodes of which the $i$'th node is darkened as a shorthand notation for external momenta corresponding to the $SU(r+1)$ fundamental weight $\omega_i$. On the right hand side, $H(SU(r+1))^{\omega_i}_Q$ is a renormalized character corresponding to the level two coset theory $H(SU(r+1))$ related to the characters presented in the introduction via the dimension
\begin{eqnarray}\label{2.4}
 \Delta=h^{w_i}_0-c/24
\end{eqnarray}
where $c$ is the $H(SU(r+1))$ coset central charge and $h^{w_i}_0$ is the fractional dimension corresponding to the coset field labeled by $(\omega_i,0)$, in general
\begin{eqnarray}\label{2.4.5}
h^{\Lambda}_{Q}=\frac{(\Lambda,\Lambda+2\rho)}{2(l+\mathfrak{g})}-\frac{(\Lambda-Q)^2}{2l}  \mod1.
\end{eqnarray}
where  $\mathfrak{g}$ denotes the dual Coxeter number while $\rho=\sum\omega_i$ is the Weyl vector.
In effort to keep our current discussion self contained let us give a short review of some results derived in \cite{gengep1}. One of the more remarking observations concerning $q$-diagrams in general is their relation to various well known identities. Indeed, studying the identities due to Jacobi and Ramanujan one finds these can be manipulated as to be exactly described by various $q$-diagrams which seem to imply these identities are only the first in an infinite such series of diagrammatic identities. In particular this observation led us to prove the complete series which in turn encapsulated all the identities needed to prove the $SO(2r)$ string function $q$-diagram correspondence for all $DHW$ of level two. Indeed, diagrammatic interpretation provide a strong tool, as such, let us recall the diagrammatic representation of the Euler identity
\begin{equation}\label{2.5}
 \includegraphics[width=0.75\linewidth,keepaspectratio=true]{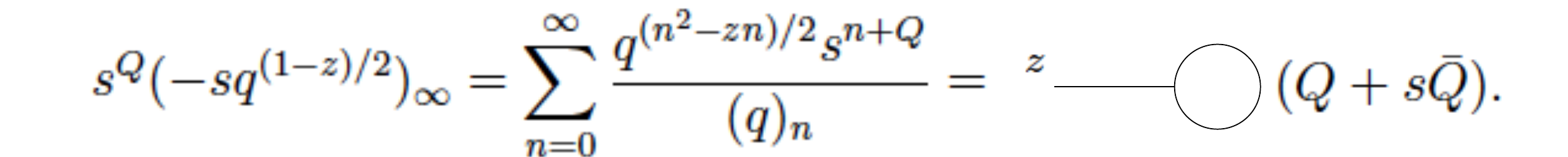}
 \end{equation}
which relates the Euler identity and the single node $q$-diagram associated with a general external momenta $z$ and the parity restriction $Q$ where $\bar{Q}=Q+1$ . As we shall find, this interpretation will come in handy when we discuss the diagrammatic interpretation of the Bailey lemma. \\
Although not trivially, we shall find the $D_r$ diagrams corresponding to the non twisted contribution are relevant to our current discussion. 
Actually, these $q$-diagrams are the diagrammatic extension of the famous Jacobi triple identity which is given the diagrammatic interpretation
\begin{equation}\label{2.6}
 \includegraphics[width=0.6\linewidth,keepaspectratio=true]{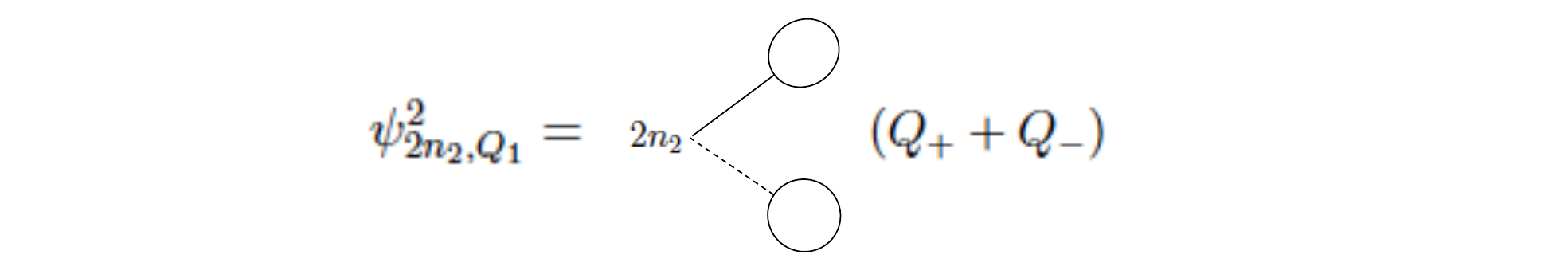}
 \end{equation}
where on the RHS $Q_+=(Q_{1},0)$, $Q_-=(1-Q_{1},1)$ and the diagram includes the so called dashed lines corresponding to negative external momenta. While on the LHS 
\begin{equation}\label{2.7}
\psi^2_{2n_2,Q_1}=(q)^{-1}_{\infty}\sum_{n_{1}=-\infty \atop n_{1} \in Z + Q_{1}/2}^{\infty}q^{2n_{1}^2-2n_{1}n_2}=\frac{1}{2}\sum_{s=\pm1}s^{Q_1}(-sq^{1/2-n_2})_{\infty}(-sq^{1/2+n_2})_{\infty}.
\end{equation}
corresponds to Jacobi theta function.
The diagrammatic extension is achieved via the study of the $D_3$ diagram with a similar arrangement of external momenta. Indeed, after some manipulation this diagram obeys the diagrammatic recursion relation
\begin{eqnarray}\label{2.8}
\includegraphics[width=0.88\linewidth,keepaspectratio=true]{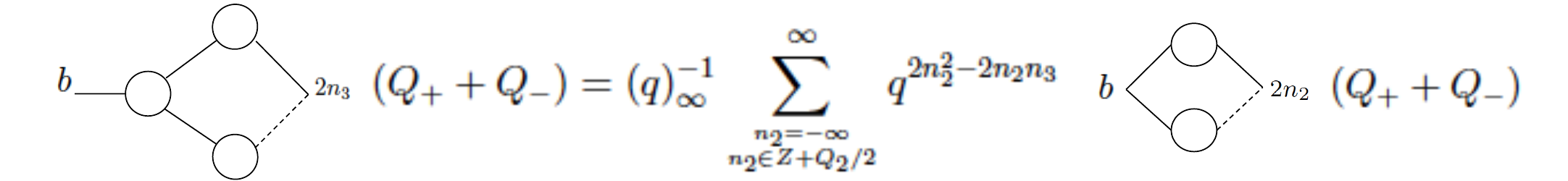}
\end{eqnarray}
which allows one to diagrammatically extend the Jacobi identity. In particular recall the $D_r(\Lambda,Q)$ diagram,
\begin{eqnarray}\label{2.9}
\includegraphics[width=0.65\linewidth,keepaspectratio=true]{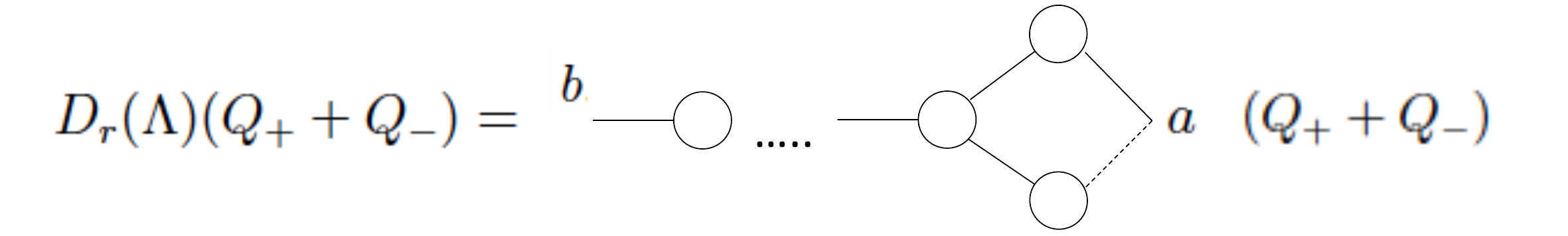}
\end{eqnarray}
where the external momenta lines are specified by an $SO(2r)$ weight $\Lambda=b\omega_1+a(\omega_{r-1}-\omega_r)$ while the summation parity restrictions are specified by the $SO(2r)$ roots $Q_+=(Q_1,...,Q_{r-1},0)$ and $Q_-=(Q_1,...,1-Q_{r-1},1)$.
In our previous paper we found this diagram to be given by,
\begin{eqnarray}\label{2.10}
D_r(\Lambda)(Q_++Q_-)=\sum^{\infty}_{\{n_i\}=-\infty    \atop n_1 \in Z, n_i \in 2Z+Q_i}
\frac{q^{\frac{1}{4}n^2-\frac{1}{2}n_1b-\frac{1}{2}n_{r-1}a}}{2(q)^{r-1}_{\infty}}\sum_{s_1=\pm1}s_1^{Q_1+n_1} 
 (-s_1q^{(1-b+n_2)/2})_b,
\end{eqnarray}
where $n=\sum n_i\alpha_i$ and $Q=\sum Q_i\alpha_i$ are roots of $SU(r)$, the summation is over $n_i$ for $i=1,...,r-1$ under the restriction $n_i=Q_i \mod2$ for $i=2,...,r-1$ and no restriction for $n_1$. On the other hand, one may expect on physical grounds that the various identities related to the twisted contributions are irrelevant.
As this is indeed the case we shall simply denote the $D_r$ diagrams according to the $SU(r)$ root $Q$ which should be understood as the combination of $Q_+$ and $Q_-$.  \\

\section{\large{The $SU(r+1)_2$ string functions as Bosonic sums}}

We mentioned in the introduction that the expression for the $SU(r+1)/U(1)^r$ coset characters, given via the ladder coset construction, implies a relation to the $SO(2r)/U(1)^r$ coset studied in $\cite{gengep1}$. Following the ladder coset construction, the $SU(r+1)/U(1)^r$ characters of eq. (\ref{1.5}) are basically given by a summing over various products of theta functions which sit on an $r$ dimensional lattice  denoted $L$  and spanned by
 \begin{equation}\label{3.1}
\epsilon_i=\sqrt{2(i+2)(i+3)}e_i,   \hspace{0.3in} e_i \cdot e_j = \delta_{ij}.
\end{equation}
with integer coefficients. Indeed, this lattice is the same as the one studied in \cite{gengep} albeit a different initial condition. To make this relation explicit, first, let us also define the dual lattice $L^{-1}$ spanned by
 \begin{equation}\label{3.2}
\zeta_i \cdot \epsilon_j = \delta_{ij},   \hspace{0.3in}              \zeta_i=\frac{1}{\sqrt{2(i+2)(i+3)}}e_i.
\end{equation}
Additionally, in what follows we find it more convenient to redefine the $k_i$ variables, appearing in the character of the product theory associated to the $A$-type generalized parafermions eq. (\ref{1.5}), such that the parity restrictions for the $m_i$'s are simply $m_i=k_i \mod2$, 
 \begin{eqnarray}\label{3.2.1}
c^{l}_{k_1,...,k_r,k_{r+1}}=\sum_{m_2,...,m_r \atop m_i= k_i \mod 2} \prod_{i=1}^{r} M(m_{i}+1,m_{i+1}+1)_i
\end{eqnarray}
note that this leaves $k_1=m_1$ and $m_{r+1}$ completely general, additionally, we define $l=m_{r+1}$ as well as $n_i=m_i+1$ for $i=1,...,r+1$.
Using these definitions the $SU(r+1)/U(1)^r$ characters can be written as a generalized theta function,
 \begin{equation}\label{3.3}
c^{l}_{k_1,...,k_r}=\eta(q)^{-r}\sum^{'}_{n_i \in k_i +1+2\mathbb{Z} }     \sum_{\{s\}}\sum_{l \in L}(-1)^{s_1+...+s_r}q^{\frac{1}{2}(l+\lambda)^2}
\end{equation}
where the summation over the $L$ lattice corresponds to $l=\sum_{i=1}^{r}l_i\epsilon_i$ and summing over all integer $l_i$ while  $s_i=0,1$ for $i=1,...,r$. Finally, the different contributions associated with the various fields appearing in the character are labeled by
 \begin{equation}\label{3.4}
\lambda(\{n^1_i\},\{s^1_i\})=\sum_{i=1}^{r}\lambda_i(n_i,n_{i+1},s_i)\zeta_i,   \hspace{0.4in}   \lambda_i(n_i,n_{i+1},s_i)=n_{i}(i+3)-n_{i+1}(i+2)(1-2s_i).
\end{equation}
where,  recall, that we have defined  $m_1=k_1$.
Next, to mimic the construction of putting $r$ single boson orbifolds at various radii on a lattice we use the beta method. Let us first introduce the character corresponding to a theory of $r$ bosons propagating on a some $r$ dimensional lattice $M$
 \begin{equation}\label{3.5}
\chi_{\Lambda,M}=\eta(q)^{-r}\sum_{m \in M}q^{\frac{1}{2}(m+\Lambda)^2}
\end{equation}
where the lattice $M$ is spanned by the original lattice $L$ and the beta vectors $\beta^i$ for $j=2,...,r$, the dual lattice $M^{-1}$ is spanned by $\omega_i$ and accordingly $m=L+\sum_{i=2}^{r}m_i\beta^i$ while $\Lambda =\sum_{i=1}^{r}\Lambda_i\omega_i$. Our objective is to match the two characters, with this in mind, consider the coset primary field associated with lowest dimension of minimal model in the product theory, i.e $n_i=k_i+1$ and $s_i=0$ for $i=1,...,r$. To include its contribution we take,
 \begin{equation}\label{3.6}
\Lambda=\lambda(\{n^1\},\{s^1\})   \equiv \lambda^{1}
\end{equation}
where $\{n^1\}=n_1,1+k_2,...,1+k_r,n_{r+1}$ and $\{s^1\}=0,...,0,s_r$ denote a specific set of values and $\lambda^{j}=\sum \lambda^{j}_i\zeta_i$ stands for $\lambda$ at the set of values specified by sets $\{n^j\}$ and $\{s^j\}$.  
Additionally, to get the correct descendent structure corresponding to the first minimal model or equivalently the trivial product theory, i.e $r=1$,  obviously we should set $\beta^1=\epsilon_1$.
To find the remaining beta vectors first set $\beta^{j}=\sum_{i=1}^r\beta^{j}_{i}\epsilon_i$ with $\beta^{j}_{i}$ non integer coefficients, $\Lambda=\lambda^1$ and $\lambda^j= \sum_i\lambda^{j}_i \frac{\epsilon_i}{\epsilon_i^2}$. Next, we demand the contribution of further $r-1$ fields are matched individually for each minimal model in the product theory,
 \begin{equation}\label{3.7}
(\Lambda+\beta^{j})^2=\lambda^{j^2}  \Rightarrow \beta^{j}_i=\frac{\lambda^j_i - \lambda^1_i}{\epsilon_i^2},
\end{equation}
where closure under the OPE of the bosonic algebra guarantees the exact matching of all other fields appearing in the character eq. (\ref{3.3}).
Solving these equations for some $j=2,...,r$ choices of $\lambda^{j}$ we find the corresponding $\beta^j$ . Clearly, these equations alone do not completely determine $\Lambda$ and $\beta$ as we only get quadratic equations. Indeed, one should complement these equations with the crucial demand that $\Lambda \in M^{-1}$ while $\beta^j \in M$,
 \begin{equation}\label{3.8}
\beta^j \cdot \beta^j \in 2\mathbb{Z},  \hspace{0.3in}  \beta^i \cdot \beta^j \in \mathbb{Z},   \hspace{0.3in}  \Lambda \cdot \beta^j \in \mathbb{Z}
\end{equation}
so that our theory will be modular invariant. As we will verify explicitly our choices for the various solutions of the quadratic equations above will satisfy these conditions. \\
To solve the beta equations consider the fields corresponding to $\{s^j\}=\{s^1\}$ for $j=2,...,r$ while $n^{j}_i=n^1_i+2\delta_{i,j}$ where $i=1,...,r+1$, note that these choices leave $n_1,n_{r+1}$ and $s_r$ completely general. Furthermore, since $n_2 \leq 3$ this choice is only possible for $k_2=0$ so that $n_2=1,3$. Let us give some explanation for this choice as it allows us to find the beta vectors for any rank and is of some physical importance. For simplicity consider the case where all $n_{i}=1$, has we have mentioned above this corresponds to demanding that the identity representative of all models are matched. Now, consider our choice for $n^j_i$ with $j>1$, here we simply take the identity field for all models in the product theory with the exception of the $j-1$ and the $j$ model which just mimics the sum over the first intermediate representation in the coset ladder representation of the $A$-type parafermion coset(see eq. \ref{3.2.1}). \\
Using eq. (\ref{3.4}) we find the corresponding $\lambda^j_i$ are given by
 \begin{equation}\label{3.9}
\lambda^j_i=\lambda^1_i+2\delta_{i,j}(i+3)-2\delta_{i+1,j}(i+2),
\end{equation}
where since $j<r+1$ we used $\delta_{i+1,j}(1-2s^1_i)=\delta_{i+1,j}$. The beta vectors are then given by eq. (\ref{3.7}),
 \begin{equation}\label{3.10}
\beta^j=\frac{1}{j+2}(\epsilon_j-\epsilon_{j-1}),
\end{equation}
for $j=2,...,r$ and $\beta^1=\epsilon_1$. Finally, to verify the first two restrictions in eq. (\ref{3.8}) define the matrix
\begin{equation}\label{3.11}
B_{ij}=\beta^i \cdot \beta^j=
2
\begin{pmatrix}
12   & -3         & .. & 0  \\
-3    &             &    &   \\
  :    &  & A_{r-1}   &  \\
 0    &             &    &   \\
\end{pmatrix}
\end{equation}
which is calculated using eqs. (\ref{3.1},\ref{3.10}). To verify the last condition we write $\Lambda$ using the dual lattice. This is done by first expressing $\epsilon_i$ via eq. (\ref{3.10}), 
 \begin{equation}\label{3.12}
\epsilon_i=\sum_{j=2}^{i}(j+2)\beta^j-\beta^1
\end{equation}
and using $\beta^j=\sum_{l=1}^rB_{jl}\omega_l$
 \begin{equation}\label{3.13}
\epsilon_i=\sum_{l=1}^r\omega_l(\sum_{j=2}^{i}(j+2)B_{jl} -B_{1l}).
\end{equation}
Next,  following the explicit form of $B$ one finds, 
 \begin{equation}\label{3.14}
\sum_{j=2}^{i}(j+2)B_{jl} -B_{1l}=2(i+3)\delta_{i,l}-2(i+2)\delta_{i,l-1}
\end{equation}
so that for $i=2,..,r$
 \begin{equation}\label{3.15}
\epsilon_i=2(i+3)\omega_i-2(i+2)\omega_{i+1}
\end{equation}
where we define $\omega_{r+1}=0$ while $\epsilon_1=24\omega_1-6\omega_2$. To express $\Lambda$ we use eq. (\ref{3.4}) 
 \begin{equation}\label{3.16}
\Lambda=\sum\lambda^{1}_i\zeta_i=\frac{1}{2}\sum_{i=1}^{r}(\frac{n^1_i}{i+2}-\frac{n^1_{i+1}(1-2s^1_{i})}{i+3})\epsilon_i
\end{equation}
which after some algebra involving eqs. (\ref{3.12} \& \ref{3.15}) is given by,
\begin{equation}\label{3.17}
\Lambda=4\omega_1n_1-\omega_2n_1+\omega_r n_{r+1}(1-2s)+\frac{1}{2}\sum_{i=2}^{r}n_i\beta^i,
\end{equation}
where from here on, to ease the discussion, we drop the $s_r$ index.
Indeed, manifestly $\Lambda \in M^{-1}_B$ and the last restriction in eq. (\ref{3.8}) is satisfied. To conclude we find the characters corresponding to the $SU(r + 1)/U(1)^r$ coset theory are equivalent to the subtraction of the two characters corresponding to two product theories of r free bosons propagating on the lattice $M_B$. \\
Finally, to facilitate calculations in the proceedings define, 
 \begin{equation}\label{3.18}
\tilde{\alpha}_i=\frac{1}{\sqrt{2}}\beta_{r+1-i}, \hspace{0.3in}  \tilde{\omega}_i=\sqrt{2}\omega_{r+1-i}
\end{equation}
so that $\tilde{\alpha}_i \cdot \tilde{\omega}_j=\delta_{ij}$ and the corresponding extension of the $SU(r)$ cartan matrix is simply $\tilde{A}_{i,j}=\frac{1}{2}B_{r+1-i,r+1-j}$.
As usual, the fundamental weights, denoted $\tilde{\omega}_i$, are defined as the dual basis
and their product is given by,
\begin{equation}\label{3.19}
\tilde{\omega}_i \cdot \tilde{\omega}_j=\tilde{A}^{-1}_{ij}.
\end{equation}
With these definitions one has,
 \begin{equation}\label{3.20}
c^{l}_{k_1,...,k_{r-2},0,k_r}=\eta(q)^{-r}\sum_{s=0,1}\sum_{m \in \tilde{n}+2M_{\tilde{A}}}(-1)^{s}q^{\frac{1}{4}(m+\tilde{\Lambda})^2}.
\end{equation}
where $\tilde{\Lambda}=\tilde{\omega}_1 n_{r+1}(1-2s) -n_1\tilde{\omega}_{r-1}+4n_1\tilde{\omega}_r$ and $\tilde{n}=\sum_{i=1}^{r-1}n_i\tilde{\alpha}_i$. \\
Finally to make the connection with the $H(D_r)$ coset consider shifting the summation by the root vector,
 \begin{equation}\label{3.21}
\tilde{v}=\sum v_i\tilde{\alpha}_i,  \hspace{0.5in} v_i=\delta_{ir}(k_r+2s-1)+1-2s+(2s-1)\sum_{j=1}^{l}\delta_{ij}(l-j)
\end{equation}
Actually, we also have $\tilde{v} \in M^{-1}_{\tilde{A}}$ 
 \begin{equation}\label{3.22}
\tilde{v}=(l+1)(2s-1)\tilde{\omega}_1+(1-2s)\tilde{\omega}_{l}+(2s-1-3k_r)\tilde{\omega}_{r-1}+(12k_r+3-6s)\tilde{\omega}_r.
\end{equation}
Additionally, when shifting $m \rightarrow m - \tilde{v}$ the resulting summation clearly depends only on $\tilde{Q}=\tilde{v}+\tilde{n} \mod2M_{\tilde{A}}$ 
 \begin{equation}\label{3.23}
\tilde{Q}=\sum \tilde{Q}_i\tilde{\alpha}_i,  \hspace{0.5in} \tilde{Q}_i=\delta_{ij}k_j-\sum_{j=1}^{l}\delta_{ij}(l-j)
\end{equation}
which is independent of $s$.
Shifting the summation we find the bosonic sum expression corresponding to characters of the $H(SU(r+1)_2)$ coset theory is given by,
\begin{eqnarray}\label{3.24}
 c^{l}_{\tilde{Q}}=\eta(q)^{-r}\sum_{s=0,1}\sum_{n \in \tilde{Q}+2M_{\tilde{A}_r}}  (-1)^s     q^{\frac{1}{4}(m-\tilde{\Lambda}_s)^2}.
\end{eqnarray}
where $\tilde{\Lambda}_s=(1-2s)\tilde{\omega}_l+(2s-2\tilde{Q}_r)\tilde{\omega}_{r-1}+(8\tilde{Q}_r-1-6s)\tilde{\omega}_r$. \\
To conclude we would like to relate these characters to the underlying $SU(r+1)$ Lie algebra. As mentioned in the introduction, the fields in our coset theory are labeled by
a level 2 dominant highest weight $\Lambda$ of the $SU(r+1)$ algebra and an element of the $SU(r+1)$ root lattice $Q$. To make the connection we simply match the corresponding fractional dimension. The fractional dimension of the coset theory primary labeled by a fundamental weight is calculated from eq. (\ref{2.4.5}) and the $SU(r+1)$ cartan matrix,
\begin{eqnarray}\label{3.25}
h^{\omega_i}_Q=\frac{i(r+1-i)}{4(r+3)}-\frac{1}{2}Q_i-\frac{1}{4}Q^2  \mod1
\end{eqnarray}
where note that here $Q \in M_A$. On the other hand, for the dimension appearing in the minimal models product theory one must add the contributions of the eta function and the  theory central charge denoted $c$,
\begin{eqnarray}\label{3.26}
d^{l}_{\tilde{Q}}=\frac{1}{4}(\tilde{\Lambda}_s+\tilde{Q})^2+\frac{c-r}{24}   \mod1
\end{eqnarray}
where $c=r(r+1)/(r+3)$. After a careful calculation one finds,
\begin{eqnarray}\label{3.27}
\frac{1}{4}\tilde{\Lambda}_s^2=\frac{l(r+1-l)}{4(r+3)}+\frac{r}{12(r+3)} +(s-\tilde{Q}_r)^2
\end{eqnarray}
let us first concentrate on $\tilde{Q}=0$.   
Collecting the contribution from the eta function and the central charge the fractional dimension,
\begin{eqnarray}\label{3.28}
d^{l}_{0}=\frac{l(r+1-l)}{4(r+3)} \mod 1,
\end{eqnarray}
along with the identification $l \leftrightarrow \omega_l$, exactly matches that of the coset theory. To complete the matching consider a general $\tilde{Q}$, the additional contributions to the fractional dimension are given by
\begin{eqnarray}\label{3.29}
\frac{1}{2}\tilde{\Lambda}_s\tilde{Q}+\frac{1}{4}\tilde{Q}^2 =\frac{1}{2}\tilde{Q}_l    -    \frac{1}{2}\tilde{Q}_r+\sum_{ij}\frac{1}{4}\tilde{Q}_i\tilde{A}_{ij}\tilde{Q}_j      \mod1.
\end{eqnarray}
Finally, following the extended cartan matrix $\tilde{A}_{ij}$ observe,
\begin{eqnarray}\label{3.30}
\frac{1}{2}\tilde{A}_{12}=\frac{1}{2}A_{12}      \mod1,  \hspace{0.2in}  \frac{1}{4}\tilde{A}_{11}=0 \mod1
\end{eqnarray}
additionally, modulo one we may use $\frac{1}{2}\tilde{Q}_r=-\frac{1}{4}A_{rr}Q_r^2$ to find 
\begin{eqnarray}\label{3.31}
\frac{1}{2}\tilde{\Lambda}_s\tilde{Q}+\frac{1}{4}\tilde{Q}^2 =\frac{1}{2}\tilde{Q}_l+\sum_{ij}\frac{1}{4}\tilde{Q}_iA_{ij}\tilde{Q}_j      \mod1
\end{eqnarray}
comparing this with the coset dimension eq. (\ref{3.25}) we can identify $\tilde{Q}_i=Q_i$ so that the fractional dimensions are in complete agreement for all fields in the theory. 
To conclude the bosonic sum representation corresponding to fundamental weights characters of the $H(SU(r+1)_2)$ coset theory labeled by the fundamental weights of $SU(r+1)$, denoted by $\omega_i$, along with an element of the $SU(r+1)$ root lattice $Q$ are given by ,
\begin{eqnarray}\label{3.32}
c^{\omega_i}_Q=\eta(q)^{-r}\sum_{s=0,1}\sum_{n \in \tilde{Q}+2M_{\tilde{A}_r}} (-1)^sq^{\frac{1}{4}(n - \tilde{\Lambda}_s)^2}
\end{eqnarray}
where $M_{\tilde{A}_r}$ denotes the root lattice corresponding to the extended $SU(r)$ matrix $\tilde{A}_r$, $\tilde{Q}=\sum Q_i \tilde{\alpha}_i$ and $\tilde{\Lambda}_s=(1-2s)\tilde{\omega}_i+(2s-2Q_r)\tilde{\omega}_{r-1}+(8Q_r-1-6s)\tilde{\omega}_r$ are a root and a weight of $\tilde{A}_r$ respectively. \\
Actually, as we shall soon find, the $SU(r+1)$ diagrams are equivalent to the level two characters only up to some dimension. With this in mind we introduce,
\begin{eqnarray}\label{3.33}
H^{\omega_i}_Q=q^{-\Delta}c^{\omega_i}_Q=(q)^{-r}_{\infty}\sum_{s=0,1}\sum_{n \in \tilde{Q}+2M_{\tilde{A}_r}} (-1)^sq^{\frac{1}{4}n^2 - \frac{1}{2} n\tilde{\Lambda}_s +d_s},
\end{eqnarray}
where $\Delta=h^{w_i}_0-c/24$ and $d_s=(Q_r-s)^2$ are some dimensions, which guarantee $H^{\omega_i}_Q$ corresponds exactly to $SU(r)$ $q$-diagrams. With this result at hand, we can now observe the so called relation to the $SO(2r)$ $q$-diagrams as well as gain some diagrammatic intuition regarding the identities needed to prove our conjecture. Indeed, note that the bosonic sum representation for the $H(SU(r+1))$ coset or equivalently the minimal models product theory contains the bosonic sum appearing in eq. (\ref{2.10}). More specifically, consider $\Lambda=0$,
\begin{eqnarray}\label{3.34}
 H^{0}_Q=\sum_{s=0,1}(q)^{-1}_{\infty} \sum_{n_r \in Q_r+2Z} q^{3n_r^2- n_r\Lambda_{s,r}/2+d_s}  (-1)^s (q)^{-r+1}_{\infty} \sum_{n \in Q+2M_{A_{r-1}}}  q^{ n^2/4  - n_{r-1} (\Lambda_{s,r-1}+3n_r)/2 }
\end{eqnarray}
One then immediately finds the following diagrammatic expression for $H^{0}_Q$,
\begin{equation}\label{3.35}
 \includegraphics[width=0.8\linewidth,keepaspectratio=true]{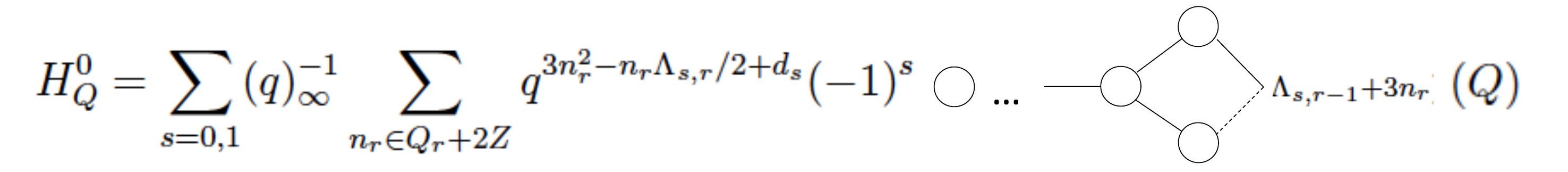}
\end{equation}
However, this is not quite the diagrammatic expression we are looking for, nonetheless, this result implies the relation, 
\begin{equation}\label{3.36}
 \includegraphics[width=0.9\linewidth,keepaspectratio=true]{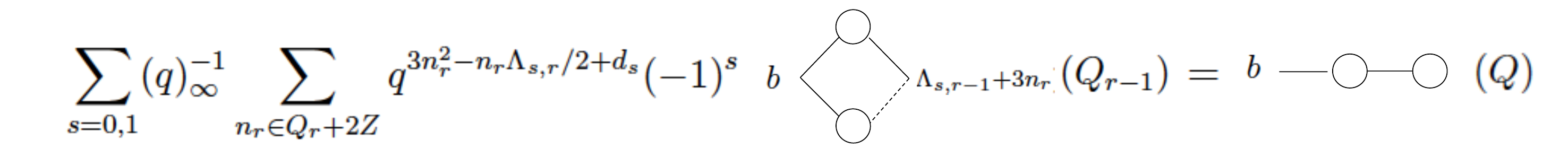}
\end{equation}
with $Q=(Q_{r-1},Q_r)$ from which our conjecture would immediately follow. How could one go about finding such a relation? Some intuition is given by examining Slater original proof \cite{sla} for the simplest identities corresponding to $SU(2)_2$,  i.e the one node diagrams $A_1(\Lambda,Q)$. The reader might recall that the one node diagram was an integral part of our work regarding the $D_r(\Lambda,Q)$ diagrams, where, using the Euler identity it was already solved. However, solving the $A_1$ diagrams, using the Euler identity, one does not find the desired expression eq. (\ref{3.33}) but an alternative expression. Indeed, in her work Slater utilises the Bailey pairs mechanism rather than the Euler identity to solve the one node diagram. Following this intuition let us first review the Bailey pairs mechanism from a diagrammatic point of view and then return to the zero momenta diagrams.

\section{\large{Bailey pairs and $q$-diagrams}}

When studying Rogers famous identities Bailey made a simple observation. Given that $\alpha_L,..,\delta_L$ are sequences satisfying,
\begin{eqnarray}\label{4.1}
 \beta_L=\sum_{r=0}^{L}\alpha_ru_{L-r}v_{L+r}   \hspace{0.4in} \text{and}   \hspace{0.4in}     \gamma_L=\sum_{r=L}^{\infty}\delta_ru_{r-L}v_{r+L}
 \end{eqnarray}
where one refers to a pair of sequences satisfying the first relation as a Bailey pair while a pair of sequences satisfying the second relation is referred to as a conjugate Bailey pair.
For such sequences one finds,
\begin{eqnarray}\label{4.2}
\sum_{L=0}^{\infty}\alpha_L \gamma_L    =\sum_{L=0}^{\infty} \beta_L\delta_L,
 \end{eqnarray}
which is a simple manner of replacing $\beta_L$ (or equivalently $\gamma_L$) and interchanging the sums. Clearly, one should impose convergence conditions for the definition of $\gamma_L$ and the interchange of sums to be meaningful. Following this observation Bailey chose $u_L=1/(q)_L$ and $v_L=1/(aq)_L$ and employed the $q$ -Saalschutz summation \cite{50y} to find,
\begin{eqnarray}\label{4.3}
\nonumber \gamma_L=\frac{(c)_L(d)_L(aq/cd)^L}{(aq/c)_L(aq/d)_L}   \frac{1}{(q)_{M-L}(aq)_{M+L}}  \\
  \delta_L= \frac{(c)_L(d)_L(aq/cd)^L}{(aq/c)_M(aq/d)_M}   \frac{(aq/cd)_{M-L}}{(q)_{M-L}}
 \end{eqnarray}
are a conjugate Bailey pair. Although Bailey considered this pair only in the $M \rightarrow \infty$ limit it was later noted by Andrews \cite{50y1,50y2} that, since $(q)_{-n} \rightarrow \infty$ for any positive $n$, by substituting this conjugate pair into eq. (\ref{4.2}) the resulting equation has the same form as the defining relation of a Bailey pair relative to $a$\footnote{Relative to $a$ meaning $\alpha(a,q)$ and $\beta(a,q)$},
\begin{eqnarray}\label{4.4}
 \sum_{L=0}^{M}   \frac{(c)_L(d)_L(aq/cd)^L}{(aq/c)_M(aq/d)_M}   \frac{(aq/cd)_{M-L}\beta_L}{(q)_{M-L}}= \sum_{L=0}^{M} \frac{(c)_L(d)_L(aq/cd)^L\alpha_L}{(aq/c)_L(aq/d)_L}   \frac{1}{(q)_{M-L}(aq)_{M+L}} 
 \end{eqnarray}
where one can identify the LHS as $\beta'_L$ and the first fraction on the RHS as $\alpha'_L$. This mechanism for producing Bailey pairs developed by Andrews has many application and is usually referred to as the Bailey chain. Following Bailey let us consider the $M \rightarrow \infty$ limit,
\begin{eqnarray}\label{4.5}
 \sum_{L=0}^{\infty}   (c)_L(d)_L(aq/cd)^L\beta_L= \frac{(aq/c)_{\infty}(aq/d)_{\infty}}{(aq/cd)_{\infty}(aq)_{\infty}} \sum_{L=0}^{{\infty}} \frac{(c)_L(d)_L(aq/cd)^L\alpha_L}{(aq/c)_L(aq/d)_L}.  
 \end{eqnarray}
This result has been extensively used by Slater to prove many of Rogers identities and has benefited both mathematicians and physicist alike. In particular we will use this result when reviewing Slater proof for the $A_1$ diagram.\\
Although not immediately apparent, let us show this relation can be given a rather intriguing diagrammatic interpretation. First, using the Pochhammer identities,
\begin{eqnarray}\label{4.6}
(c)_n=\frac{(-c)^nq^{n(n-1)/2} }{(q/c)_{-n}}   \hspace{0.4in}  \text{ and }  \hspace{0.4in}  (c)_n=\frac{(c)_{\infty}}{(cq^n)_{\infty}}
 \end{eqnarray}
 to replace all the finite Pochhammer symbols on both sides,
\begin{eqnarray}\label{4.7}
\nonumber (aq/cd)_{\infty} \sum_{L=0}^{\infty}   (q^{1-L}/c)_{\infty}(q^{1-L}/d)_{\infty}a^Lq^{L^2}\beta_L=     \hspace{2.8in} \\
 \hspace{0.3in}  \frac{1}{(aq)_{\infty}} \sum_{L=0}^{\infty} (aq^{1+L}/c)_{\infty}(aq^{1+L}/d)_{\infty}(q^{1-L}/c)_{\infty}(q^{1-L}/d)_{\infty}a^Lq^{L^2}\alpha_L   
\end{eqnarray}
Next, for our purposes , consider $a=1,q$ so that one can replace $a \rightarrow q^{a}$ where $a=0,1$. Additionally, with no loss of generality, replace $c \rightarrow-s_cq^{(1+c+a)/2}$ where $s_c=\pm1$ and similarly for $d \rightarrow-s_dq^{(1+d+a)/2}$. These redefinitions are motivated by the following observation. Consider, for example, the first Pochhammer on the LHS using eq. (\ref{2.5}),
 \begin{equation}\label{4.8}
 \includegraphics[width=0.8\linewidth,keepaspectratio=true]{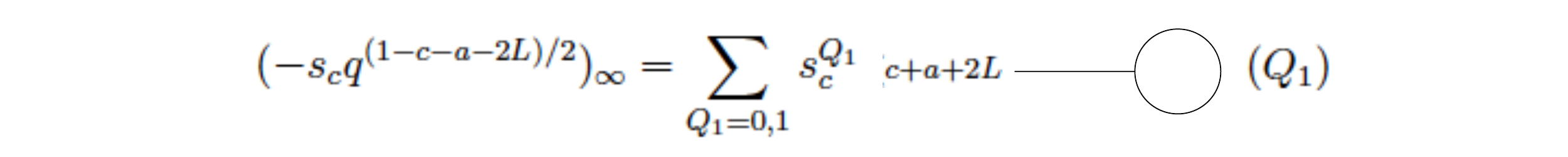}
\end{equation}
Replacing all the $L$ dependent Pochhammer symbols in eq. (\ref{4.7}) we find,
 \begin{equation}\label{4.9}
 \includegraphics[width=0.8\linewidth,keepaspectratio=true]{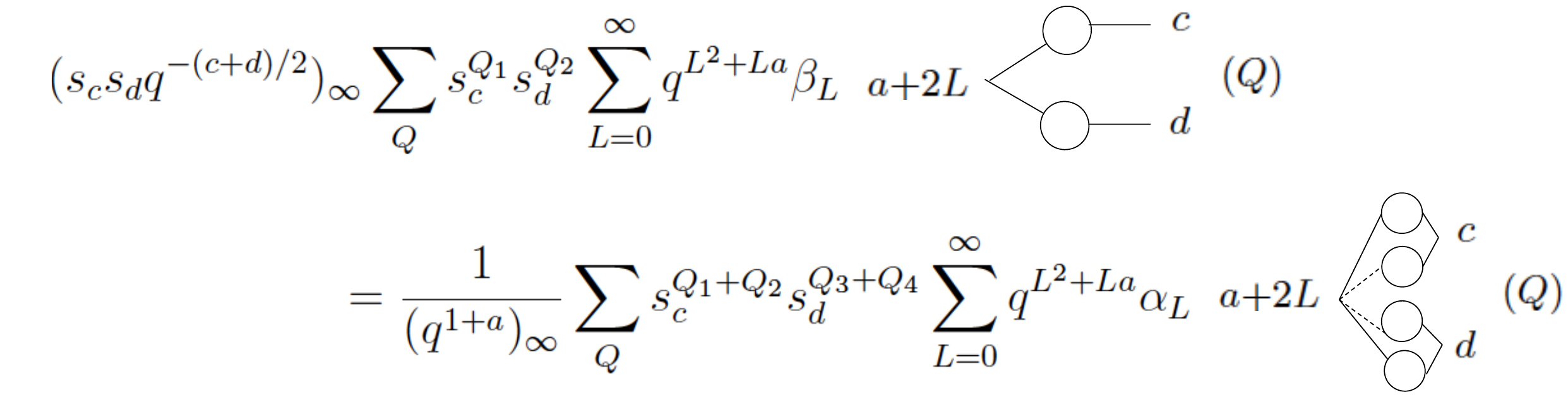}
\end{equation}
where we label the nodes from top to bottom, $Q=(Q_1,..,Q_r)$ such that $r$ is the length of the associated diagram and the summation is over $Q_i=0,1$ for $i=1,...,r$, i.e all possible parity restrictions. Having found a diagrammatic interpretation for eq. (\ref{4.7}) one can write down a diagrammatic identity for any Bailey pair. Furthermore, recognising the $D_2$ diagrams one may use the $SO(2r)$ diagrammatic recursion relation, found in our previous paper \cite{gengep1}, to find an infinite number of new identities relating Bailey pairs or equivalently an infinite number of new conjugate Bailey pairs. Finally, although we have assumed $a=1,q$ this is just a matter of convenience and one can easily consider any $a$.\\
For the purpose of proving our conjecture it is enough to consider the diagrammatic Bailey pair relation (eq. \ref{4.9}) in the $d \rightarrow -\infty$ limit,
 \begin{equation}\label{4.10}
 \includegraphics[width=0.8\linewidth,keepaspectratio=true]{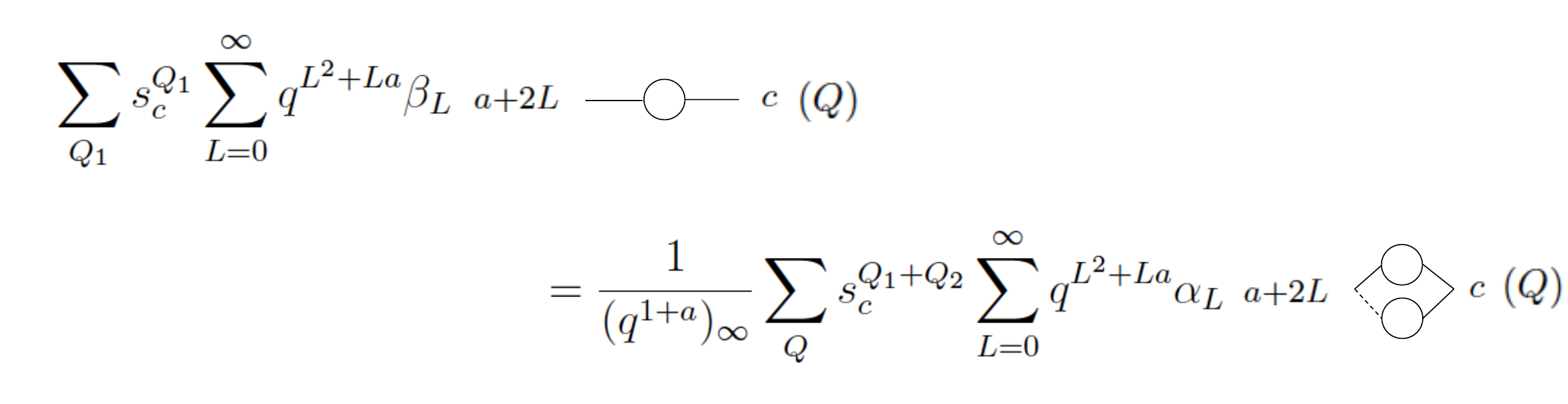}
\end{equation}
This expression can be simplified by summing over $\sum_{s_c=\pm1}s_c^{Q'_1}$ to find,
 \begin{equation}\label{4.11}
 \includegraphics[width=0.8\linewidth,keepaspectratio=true]{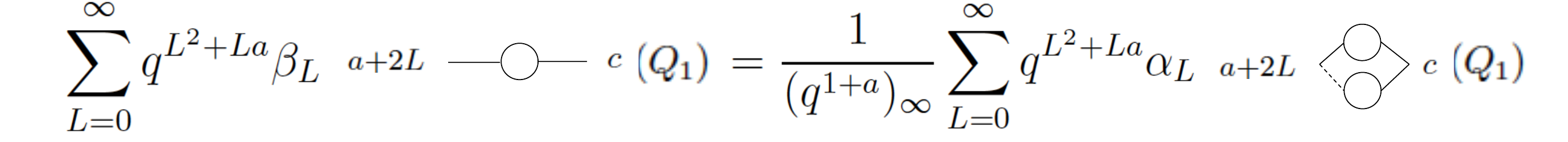}
\end{equation}
where we relabel $Q'_1 \rightarrow Q_1$ while as explained in section $2$ the $D_2$ diagram on the RHS contains two nodes and $Q_1$ should be understood as $Q_++Q_-$. At this point the reader might note the similarities between this relation and eq. (\ref{3.36}). Indeed, in the next section, we find an appropriate Bailey pair to prove eq. (\ref{3.36}) and subsequently our conjecture for the zero momenta diagrams.

\section{\large{The zero momenta diagrams}  }

Rogers identities for the characters of the $H(SU(2)_2)$ coset theory, better known as the first Minimal model $M(3,4)$ or the Ising model, relate these characters with the $A_1(\Lambda,Q)$ diagrams. In particular the characters of the Ising model are given by eq. (\ref{3.33}),
\begin{eqnarray}\label{5.1}
 H^{0}_Q=(q)^{-1}_{\infty}\sum_{s=0,1}\sum_{n \in Q+2Z} q^{3n^2 - n \cdot \Lambda_s/2+d_s}(-1)^s
\end{eqnarray}
where,
\begin{eqnarray}\label{5.2}
\Lambda_s=8Q-1-6s   \hspace{0.4in} \text{ and }  \hspace{0.4in} 
d_s=(Q-s)^2
\end{eqnarray}
Slater \cite{ber} obtained the corresponding Rogers identities by using the following Bailey pairs,
\begin{equation}\label{5.3}
\begin{array}{llllll} \hline\hline
   \beta_n & \alpha_0 & \alpha_{3n-1} & \alpha_{3n} & \alpha_{3n+1} & a \\ \hline 
  q^{n^2}/(q)_{2n} & 1 & -q^{3n^2-n}  & q^{3n^2-n}+q^{3n^2+n}  & -q^{3n^2+n} & 0 \\ \hline
  q^{n^2+n}/(q^2)_{2n} & 1  & q^{3n^2-2n}   & q^{3n^2+2n}  & -q^{3n^2+4n+1}-q^{3n^2+2n} & 1  \\
  \hline\hline
\end{array}
 \end{equation}
where, when using these pairs, one should be mindful as to set $\alpha_0=1$ for both cases.
Additionally, Slater considered eq. (\ref{4.11}) in the $c\rightarrow \infty$ limit,
 \begin{eqnarray}\label{5.4}
 \sum_{L=0}^{\infty}q^{L^2+La}\beta_L = \frac{1}{(q^{1+a})_{\infty}}\sum_{L=0}^{\infty}q^{L^2+La}\alpha_L.
 \end{eqnarray}
Indeed, putting the Bailey pairs of eq. (\ref{5.3}) into this relation one finds (after a bit of algebra),
 \begin{equation}\label{5.5}
 \includegraphics[width=0.5\linewidth,keepaspectratio=true]{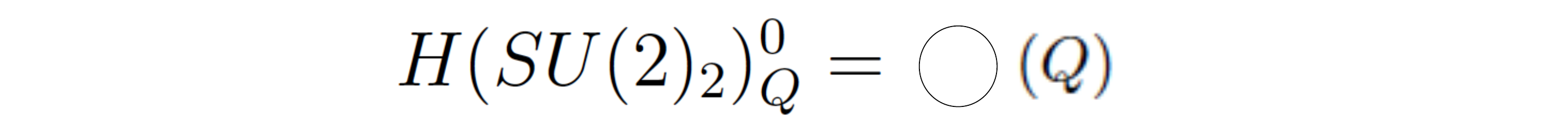}
\end{equation}
Which clearly agrees with our conjecture for the $H(SU(2)_2)$ characters. To extend this result and prove our conjecture one simply needs to consider the Bailey pairs of eq. (\ref{5.3}) for a general $c$ as is implied by the diagrammatic form of eq. (\ref{4.11}). More specifically, note that generally $\beta_L(a)$ can written as,
 \begin{eqnarray}\label{5.6}
\beta_L = \frac{q^{L^2+La}}{(q^{1+a})_{2L }}. 
  \end{eqnarray}
Then one finds,
 \begin{eqnarray}\label{5.7}
q^{L^2+La}\beta_L = \frac{q^{2L^2+2La}}{(q^{1+a})_{2L }} 
  \end{eqnarray}
which can be put in a diagrammatic form by simply redefining $a+2L = b$, using
\begin{eqnarray}\label{5.8}
(q)_{n+a}=(q)_a(q^{1+a})_n
 \end{eqnarray}
and noting that $a^2=a$ to find,
\begin{eqnarray}\label{5.9}
q^{L^2+La}\beta_L =q^{-a/2}(q)_{a}\frac{q^{b^2/2}}{(q)_{b} }(1+(-1)^{b+a}) 
 \end{eqnarray}
which, following our diagrammatic rules, when summed over $b$ exactly produces one node. 
Subsequently the LHS of eq. (\ref{4.11}) is given by,
\begin{equation}\label{5.10}
 \includegraphics[width=\linewidth,keepaspectratio=true]{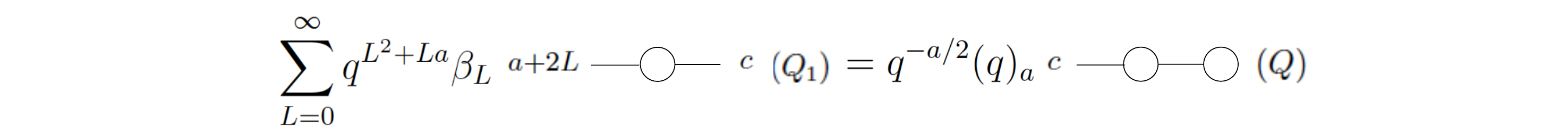}
\end{equation}
with $Q=(Q_1,a)$ which up to a multiplicative factor exactly matches the expected relation eq. (\ref{3.36}) with $a=Q_2$. The RHS is a bit more tedious, let us first denote
\begin{equation}\label{5.11}
 \includegraphics[width=0.33\linewidth,keepaspectratio=true]{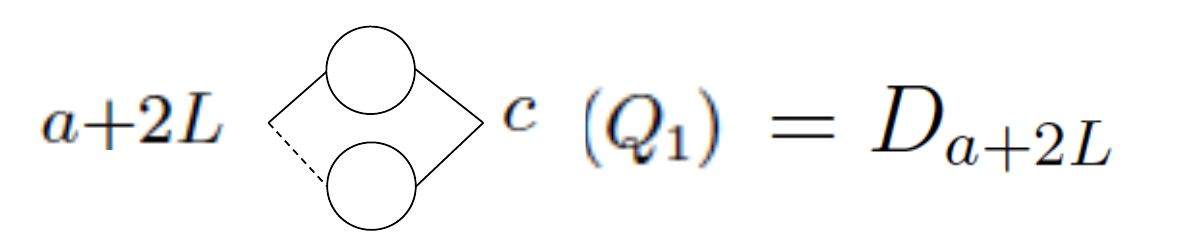}
\end{equation}
where $D_{a}=D_{-a}$ which amounts to exchanging the two nodes and noting that the combination of $Q_++Q_-$ is symmetric under this exchange.  Replacing $L=3n+j$ and using the notation introduced above,
 \begin{eqnarray}\label{5.12}
\sum_{L=0}^{\infty}q^{L^2+La} \alpha_L D_{a+2L}=\sum_{n=0}^{\infty}\sum_{j=-1}^{1}q^{9n^2+3n(2j+a)+j(j+a)}\alpha_{3n+j}D_{a+6n+2j}.
\end{eqnarray}
where $\alpha(a)_{-1}=0$.
To proceed recall $\alpha(a)_0=1$ while $\alpha_L(a)$, where $a=0,1$ and $L > 0$, can be written as,
\begin{eqnarray}\label{5.13}
\alpha_L(a)=\begin{cases} (-1)^{1+a}q^{3n^2-n(1-3a)} & L=3n-1+a \\ 
(-1)^{a}q^{3n^2}(q^{n(5a-1)+a}+q^{n(1+a)}) & L=3n+a \\
(-1)^{1+a}q^{3n^2+n(1-3a)} & L=3n+1-2a
\end{cases}
  \end{eqnarray}
and consider the contribution of $j=a$
 \begin{eqnarray}\label{5.14}
\sum_{n=0}^{\infty}q^{9n^2+9na+2a}\alpha_{3n+a}D_{3a+6n}=\sum_{n=1-a}^{\infty}D_{3a+6n}(-1)^{a}q^{12n^2+n(14a-1)+3a} \hspace{0.8in}\\
\nonumber +\sum_{n=0}^{\infty}D_{3a+6n}(-1)^{a}q^{12n^2+n(1+10a)+2a}
  \end{eqnarray}
where we separate the sums to keep $\alpha_0=1$.
At this point, one may note, that the sums appearing here range over positive $n$ while the sum appearing on the LHS of eq. (\ref{3.36}) ranges over both negative and positive $n$. This motivates substituting $n \rightarrow -n-a$ for the first sum,
  \begin{eqnarray}\label{5.15}
\sum_{n=1-a}^{\infty}D_{3a+6n}(-1)^{a}q^{12n^2+n(14a-1)+3a}=\sum_{n=-\infty}^{-1}D_{3a+6n}(-1)^{a}q^{12n^2+n(1+10a)+2a}
  \end{eqnarray}
where we have used the diagram symmetry $D_{-a}=D_{a}$. Indeed, combining the two sums we find,
 \begin{eqnarray}\label{5.16}
\sum_{n=0}^{\infty}q^{9n^2+9na+2a}\alpha_{3n+a}D_{3a+6n}=\sum_{n=-\infty}^{\infty}(-1)^{a}q^{12n^2+n(1+10a)+2a}D_{3a+6n}.
  \end{eqnarray}
Next, consider the contribution of $j=a-1,1-2a$,
 \begin{eqnarray}\label{5.17}
\nonumber \sum_{n=1-a}^{\infty}q^{9n^2+1-a+3n(3a-2)}\alpha_{3n+a-1}D_{6n+3a-2}+\sum_{n=a}^{\infty}q^{9n^2+1-a+3n(2-3a)}\alpha_{3n+1-2a}D_{6n-3a+2}   \hspace{1in}\\
=(-1)^{a+1}q^{1-a}\Big(\sum_{n=1-a}^{\infty}q^{12n^2+n(12a-7)}D_{6n+3a-2}+\sum_{n=a}^{\infty}q^{12n^2+n(7-12a)}D_{6n-3a+2}\Big). \hspace{0.3in}
\end{eqnarray}
As for the $j=a$ contribution, this can be written as one sum ranging over both positive and negative integers,
 \begin{eqnarray}\label{5.18}
\nonumber \sum_{n=1-a}^{\infty}q^{9n^2+1-a+3n(3a-2)}\alpha_{3n+a-1}D_{6n+3a-2}+\sum_{n=a}^{\infty}q^{9n^2+1-a+3n(2-3a)}\alpha_{3n+1-2a}D_{6n-3a+2}   \hspace{1in}\\
=\sum_{n=-\infty}^{\infty}(-1)^{a+1}q^{12n^2+n(7-2a)+1-a}D_{6n-a+2}  \hspace{0.5in}
\end{eqnarray}
by changing $n \rightarrow (2a-1)n$ at the first sum while changing $n \rightarrow (1-2a)n$ at the second sum and using the diagram symmetry.
Finally, collecting these results one finds,
 \begin{eqnarray}\label{5.19}
\frac{1}{(q^{1+a})_{\infty}}\sum_{L=0}^{\infty}q^{L^2+La} \alpha_L D_{a+2L}=\frac{(q)_{a}}{(q)_{\infty}}\sum_{s=0,1}\sum_{n=-\infty}^{\infty}(-1)^sq^{12n^2+n(1+6s+4a)+s(1+a)}D_{6n+a+2s}
\end{eqnarray}
Indeed, by changing $2n+a \rightarrow n$ so that $n \in a+2Z$, this matches the RHS of eq. (\ref{3.36})
\begin{equation}\label{5.20}
 \includegraphics[width=\linewidth,keepaspectratio=true]{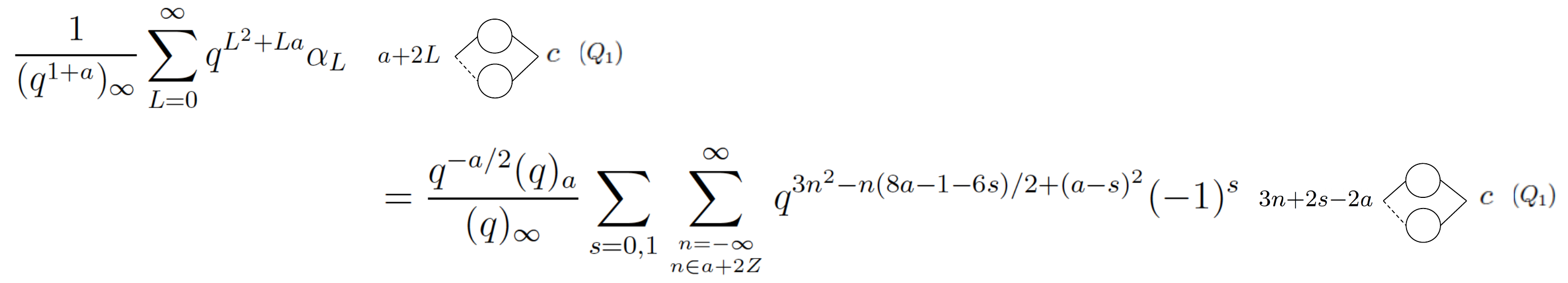}
\end{equation}
again up to $q^{-a/2}(q)_{a}$ and $a=Q_2$. Thus, equating the two sides (eqs. \ref{5.10} \& \ref{5.20}) we find the expected relation eq. (\ref{3.36}). As discussed in section ($3$), using this relation in eq. (\ref{3.35}) proves the conjectured identity for the zero momenta diagrams,
\begin{equation}\label{5.21}
 \includegraphics[width=0.58\linewidth,keepaspectratio=true]{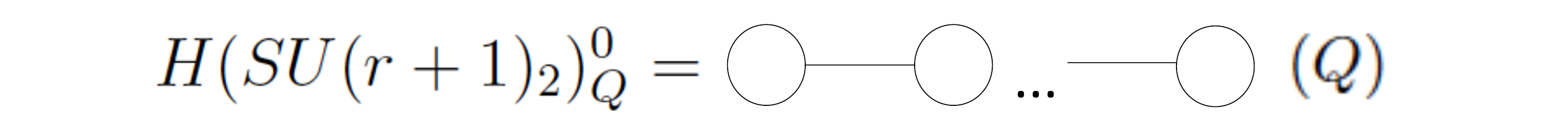}
\end{equation}
Before we proceed consider the non zero momenta characters,
\begin{eqnarray}\label{5.22}
 H^{\omega_i}_Q=\sum_{s=0,1}(q)^{-r}_{\infty} \sum_{n_r \in Q_r+2Z} q^{3n_r^2- n_r\Lambda_{s,r}/2+d_s}  (-1)^s  \sum_{n \in Q+2M_{A_{r-1}}}  q^{ n^2/4 -n_i\Lambda_{s,i}/2 - n_{r-1} (\Lambda_{s,r-1}+3n_r)/2 }
\end{eqnarray}
As discussed in our previous paper, for any weight belonging to the weight lattice of $SU(r)$ one can define
\begin{eqnarray}\label{5.23}
\Lambda_{s,i}\omega_i + (\Lambda_{s,r-1}+3n_r)\omega_{r-1}=n-a\omega_{r-1}
\end{eqnarray}
where $n$ is a root of $SU(r)$. Thus, the inner sum appearing here can be replaced with $D_r(a\omega_{-},Q+n)$ i.e the zero momenta diagram with some external line. However, contrary to the zero momenta diagrams, we expect to find the $A_r(\omega_i,Q)$ diagram for which the $i$'th node is black. Clearly, using the zero momenta $D_r$ diagram does not produce black nodes. Instead, let us consider the $A_r$ diagrams corresponding to $\Lambda=\omega_i$, for $i=1,...,r-2$. These diagrams clearly include the $A_2(b\omega_1,Q)$ diagrams and one can apply eq. (\ref{3.36}) to find,
\begin{eqnarray}\label{5.24}
 \includegraphics[width=0.9\linewidth,keepaspectratio=true]{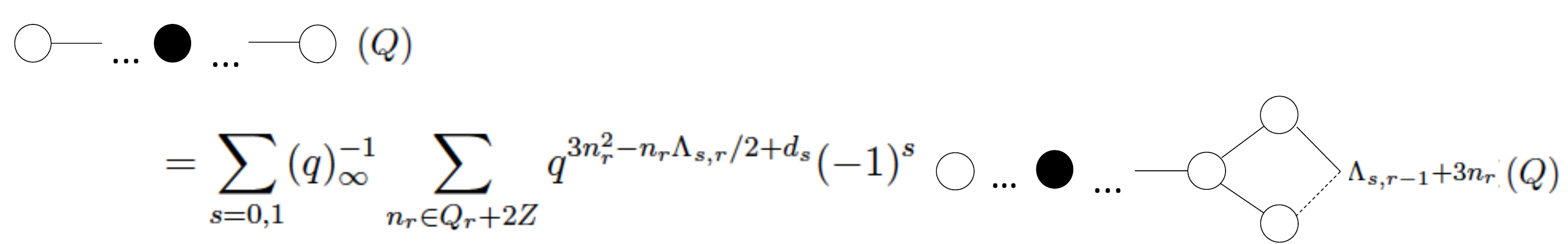}
\end{eqnarray}
where both diagrams contain $r$ nodes of which the $i$'th node is black. Additionally, following the convention set in section $2$ the $Q$ vector associated with the $D_r$ diagram is given by $Q=(Q_1,...,Q_{r-1})$ and should be understood as the combination $Q_++Q_-$.
This relation leads to the conclusion that to prove our conjecture for the $A_r(\omega_i,Q)$ diagrams one should find a bosonic sum representation for the $D_r(\omega_i+a\omega_{-},Q)$ diagrams. Accordingly, in the following section, we first study these diagrams which were not studied in our paper regarding the $D_r$ diagrams.

\section{\large{The $D_r$ fundamental weights diagrams}  }

Consider the $D_r$ diagrams appearing in eq. (\ref{5.24}), clearly for any $i<r-2$ one can use the diagrammatic recursion relation eq. (\ref{2.8}). However, eventually one comes across the following $D_3$ diagram,
\begin{eqnarray}\label{6.1}
 \includegraphics[width=0.55\linewidth,keepaspectratio=true]{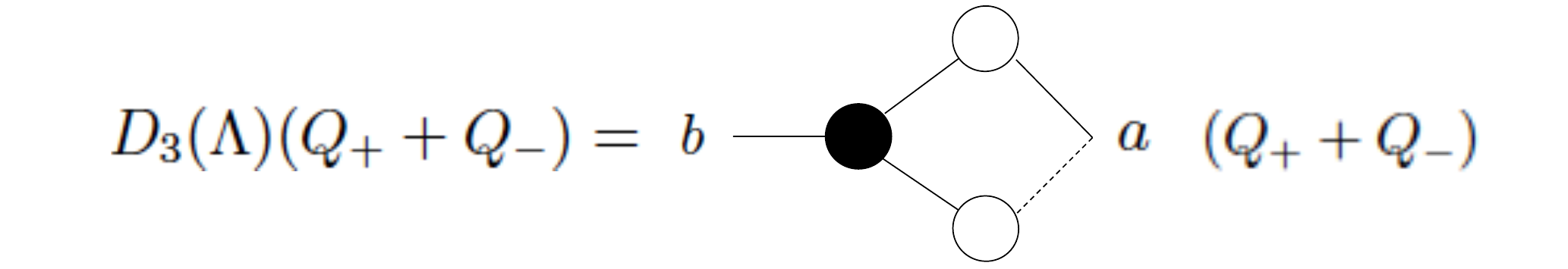}
\end{eqnarray}
where $\Lambda=(1+b)\omega_1+a\omega_{-}$. Actually, one may try to continue the reduction via the diagrammatic recursion relation but this procedure will introduce new diagrams, specifically the $D_3(b\omega_1+a\omega_{-}+\omega_+)$ diagram. Instead, let us take a closer look at the bosonic sum representation given by eq. (\ref{2.10})
\begin{eqnarray}\label{6.2}
D_3(\Lambda,Q)=\sum^{\infty}_{\{n_i\}=-\infty    \atop n_1 \in Z, n_2 \in 2Z+Q_2}
\frac{q^{\frac{1}{4}n^2-\frac{1}{2}n_1(1+b)-\frac{1}{2}n_{2}a}}{2(q)^{2}_{\infty}}\sum_{s_1=\pm1}s_1^{Q_1+n_1}
 (-s_1q^{(1-b-1+n_2)/2})_{b+1}.
\end{eqnarray}
To avoid running into new diagrams, one may try to write this diagram in terms of $D_3$ diagrams with external lines specified by a weight of the form $b\omega_1+c\omega_{-}$ with some $c$.
For this purpose shift $n=l-\alpha_2$,
\begin{eqnarray}\label{6.3}
D_3(\Lambda,Q)=\sum^{\infty}_{\{n_i\}=-\infty    \atop n_1 \in Z, n_2 \in 2Z+Q_2+1}
\frac{q^{\frac{1}{4}n^2-\frac{1}{2}n_1b-\frac{1}{2}n_{2}(a+2)+\frac{1}{2}(1+a)}}{2(q)^{2}_{\infty}}\sum_{s_1=\pm1}s_1^{Q_1+n_1}
 (-s_1q^{(1-b+n_2)/2-1})_{b+1}.
\end{eqnarray}
To proceed note that following the Pochhammer symbol definition,
\begin{eqnarray}\label{6.4}
 (-s_1q^{(1-b+n_2)/2-1})_{b+1}= (-s_1q^{(1-b+n_2)/2})_{b}(1+s_1q^{(n_2-1-b)/2}).
\end{eqnarray}
Indeed, replacing the Pochhammer symbol with this expression we find,
\begin{eqnarray}\label{6.5}
 \includegraphics[width=0.9\linewidth,keepaspectratio=true]{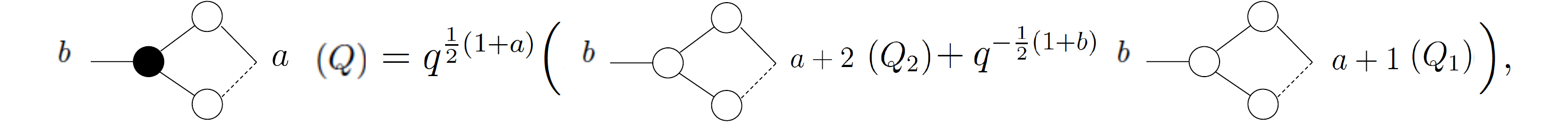}
\end{eqnarray}
where $Q_1=Q+\alpha_1+\alpha_2$ and $Q_2=Q+\alpha_2$. If we regard the $D_3$ diagram as the tail of some $D_r(a\omega_{-}+\omega_{r-2})$, i.e $b=b_{r-3}$, we find the following diagrammatic relation
\begin{eqnarray}\label{6.6}
 \includegraphics[width=0.9\linewidth,keepaspectratio=true]{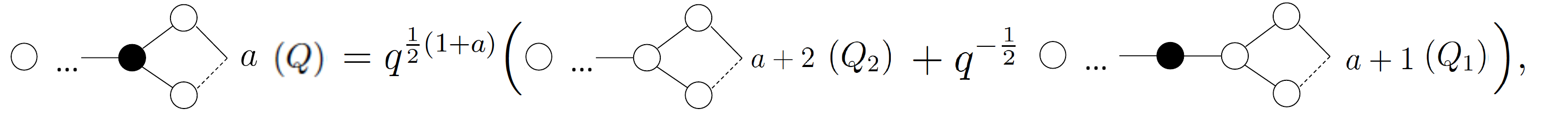}
\end{eqnarray}
where all diagrams contain $r$ nodes while $Q_1=Q+\alpha_{r-2}+\alpha_{r-1}$ and $Q_2=Q+\alpha_{r-1}$. A bosonic sum representation for the first diagram that appears on the RHS is then given by eq. (\ref{2.10})
\begin{eqnarray}\label{6.7}
D_r\big((a+2)\omega_{-},Q_2\big)=
\sum^{\infty}_{\{n_i\}=-\infty    \atop n \in Q_2+2M_{A_{r-1}}}
\frac{q^{\frac{1}{4}n^2-\frac{1}{2}n_{r-1}(a+2)}}{(q)^{r-1}_{\infty}},
\end{eqnarray}
additionally, shifting $n=l+\alpha_{r-1}$ we find
\begin{eqnarray}\label{6.8}
D_r\big((a+2)\omega_{-},Q_2\big)=
\sum^{\infty}_{\{n_i\}=-\infty    \atop n \in Q+2M_{A_{r-1}}}
\frac{q^{\frac{1}{4}n^2-\frac{1}{2}n_{r-2}-\frac{1}{2}n_{r-1}a-\frac{1}{2}(a+1)}}{(q)^{r-1}_{\infty}}.
\end{eqnarray}
Finally, using this result in eq. (\ref{6.6}),
\begin{eqnarray}\label{6.9}
D_r(\omega_{r-2}+a\omega_{-},Q)=\sum^{\infty}_{\{n_i\}=-\infty    \atop n \in Q+2M_{A_{r-1}}}
\frac{q^{\frac{1}{4}n^2-\frac{1}{2}n_{r-2}-\frac{1}{2}n_{r-1}a}}{(q)^{r-1}_{\infty}}
+q^{a/2}D_r\big(\omega_{r-3}+(a+1)\omega_{-},Q_1\big),
\end{eqnarray}
so that the diagram corresponding to $\Lambda=\omega_{r-2}+a\omega_{-}$ and $Q$ is given in terms of a bosonic sum and the diagram corresponding to $\omega_{r-3}+(a+1)\omega_{-}$ and $Q_1$. \\
To find a similar relation for $\Lambda=\omega_{i}+a\omega_{-}$ with $i=1,...,r-2$, consider eq. (\ref{6.9}) summed over $a$ in the following manner
\begin{eqnarray}\label{6.10}
(q)_{\infty}^{-1}\sum_{a \in Q_r+2Z}q^{a(a-c)/2}.
\end{eqnarray}
The sum appearing on the LHS then exactly matches the recursion relation eq. (\ref{2.8}),
\begin{eqnarray}\label{6.11}
(q)_{\infty}^{-1}\sum_{a \in Q_r+2Z}q^{a(a-c)/2}D_r(\omega_{r-2}+a\omega_{-},Q)=D_{r+1}(\omega_{r-2}+c\omega_{-},Q)
\end{eqnarray}
with $Q=(Q_1,...,Q_{r-1},Q_r)$. For the RHS the first term is easily written as,
\begin{eqnarray}\label{6.12}
(q)_{\infty}^{-1}\sum_{a \in Q_r+2Z}q^{a(a-c)/2}\sum^{\infty}_{\{n_i\}=-\infty    \atop n \in Q+2M_{A_{r-1}}}
\frac{q^{\frac{1}{4}n^2-\frac{1}{2}n_{r-2}-\frac{1}{2}n_{r-1}a}}{(q)^{r-1}_{\infty}}
=\sum^{\infty}_{\{n_i\}=-\infty    \atop n \in Q+2M_{A_{r}}}
\frac{q^{\frac{1}{4}n^2-\frac{1}{2}n_{r-2}-\frac{1}{2}n_{r}c}}{(q)^{r}_{\infty}}.
\end{eqnarray}
Finally, for the second sum shift $l=a+1$,
\begin{eqnarray}\label{6.13}
\sum_{a \in Q_r+2Z}\frac{q^{a(a-c+1)/2}}{(q)_{\infty}}D_r\big(\omega_{r-3}+(a+1)\omega_{-},Q_2\big)=
\sum_{l \in Q_r+1+2Z}\frac{q^{l(l-1-c)/2+c/2}}{(q)_{\infty}}D_r\big(\omega_{r-3}+l\omega_{-},Q_2\big)
\end{eqnarray}
The resulting sum matches the recursion relation, furthermore the solution is given by,
\begin{eqnarray}\label{6.14}
\sum_{a \in Q_r+2Z}\frac{q^{a(a-c+1)/2}}{(q)_{\infty}}D_r\big(\omega_{r-3}+(a+1)\omega_{-},Q_2\big)=
q^{c/2}D_{r+1}\big(\omega_{r-3}+(c+1)\omega_{-},Q_2\big)
\end{eqnarray}
which is the same expression with $a$ replaced with $c$, an additional node in the diagram and $Q_2=Q+\alpha_{r-2}+\alpha_{r-1}+\alpha_{r}$. To conclude we find,
\begin{eqnarray}\label{6.15}
D_{r+1}(\omega_{r-2}+c\omega_{-},Q)=\sum^{\infty}_{\{n_i\}=-\infty    \atop n \in Q+2M_{A_{r}}}
\frac{q^{\frac{1}{4}n^2-\frac{1}{2}n_{r-2}-\frac{1}{2}n_{r}c}}{(q)^{r}_{\infty}}
+q^{c/2}D_{r+1}\big(\omega_{r-3}+(c+1)\omega_{-},Q_2\big),
\end{eqnarray}
which is clearly viable for any $r$ thus effectively moves the black nodes one node back. As this process can be repeated we find the following diagrammatic relation,
\begin{eqnarray}\label{6.16}
 \includegraphics[width=0.95\linewidth,keepaspectratio=true]{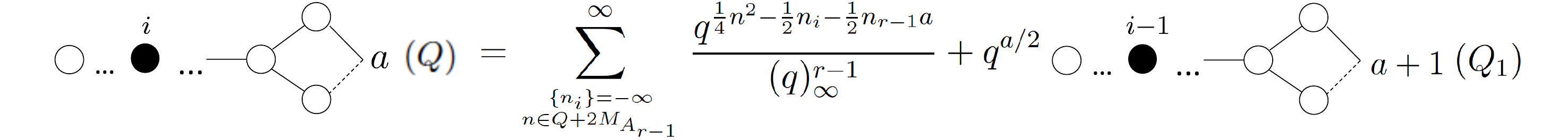}
\end{eqnarray}
with $Q$ some $SU(r)$ root while $Q_1=Q+\sum_{l=i}^{r-1}\alpha_l$, which is a recursive relation relating diagrams of $\Lambda=\omega_i+a\omega_{-}$ with diagrams of $\Lambda=\omega_{i-1}+(a+1)\omega_{-}$. Since we have a bosonic sum representation for the $i=0$ diagram $D_r(a\omega_{-},Q)$ we can apply this relation repeatedly to find a bosonic sum representation for any $i=0,...,r-2$,
\begin{eqnarray}\label{6.17}
 \includegraphics[width=0.9\linewidth,keepaspectratio=true]{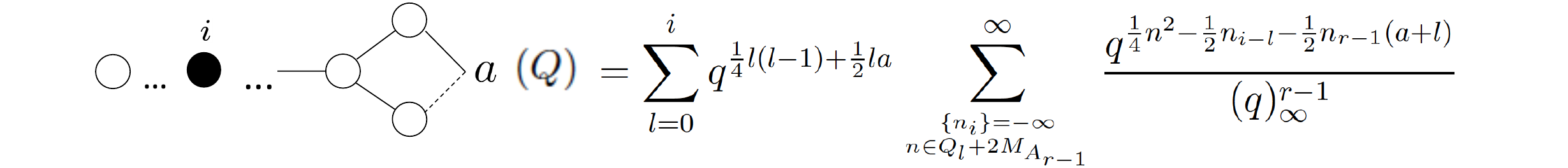}
\end{eqnarray}
where $n_0=0$ and $Q_l$ is defined by the recursion relation,
\begin{eqnarray}\label{6.18}
Q_{l+1}=Q_{l}+\sum_{j=i-l}^{r-1}\alpha_j
\end{eqnarray}
and the initial condition $Q_0=Q$. We have thus found an infinite number of new $q$ sums identities corresponding to bosonic sum representations for any $D_r(\omega_i+a\omega_{-},Q)$ diagram for $i=0,1,...,r-2$. As discussed in the previous section, these identities are intimately related to the $A_r(\omega_i,Q)$ diagrams. Indeed, these identities will furnish a way to prove the conjectured identities for the $A_r(\omega_i,Q)$ diagrams which are the subject of the next section.

\section{\large{The $A_r$ non zero momenta diagrams}  }

Following up on our discussion in the end of section ($5$), one can use the bosonic sum description for the $D_r$ diagrams to find a bosonic sum for any of the $A_r$ diagrams. More specifically, recall eq. (\ref{5.24}) which we reproduce for convenience,
\begin{eqnarray}\label{7.1}
 \includegraphics[width=0.9\linewidth,keepaspectratio=true]{15s}
\end{eqnarray}
where,
\begin{eqnarray}\label{7.2}
\Lambda_{s,r-1}=2s-2Q_r, \hspace{0.3in}
\Lambda_{s,r}=(8Q_r-1-6s),  \hspace{0.3in}   d_s=Q_r(1-2s)+s.
\end{eqnarray}
To replace the $D_r$ diagram with the bosonic description of eq. (\ref{6.17}) simply take $a=\Lambda_{s,r-1}+3n_r$ to find, 
\begin{eqnarray}\label{7.3}
A_r(\omega_i,Q)=(q)^{-r}_{\infty}\sum_{s=0,1} \sum_{l=0}^{i}   (-1)^s\sum^{\infty}_{\{n_i\}=-\infty    \atop n \in \tilde{Q}+2M_{\tilde{A}_r}}
q^{\frac{1}{4}n^2-\frac{1}{2}n\tilde{\lambda}_{s,l}+h_{s,l}}
\end{eqnarray}
where $h_{s,l}$ is some dimension, $\lambda_{s,l}$ is a weight and  $\tilde{Q}$ is a root of $\tilde{A}_r$ defined as,
\begin{eqnarray}\label{7.4}
\lambda_{s,l}=\tilde{\omega}_{i-l}+\tilde{\omega}_{r-1}(\Lambda_{s,r-1}+l)+\tilde{\omega}_{r}(\Lambda_{s,r}-3l),  \\
\nonumber h_{s,l}=d_s+\frac{1}{2}l\Lambda_{s,r-1}+\frac{1}{4}l(l-1), \hspace{0.91in} \\
\nonumber  \tilde{Q}_l=\sum^{r-1}_{i=1}Q_{l,i}\tilde{\alpha}_i+Q_r\tilde{\alpha}_r.   \hspace{1.5in}
\end{eqnarray}
Indeed we find that all the $A_r$ diagrams can be written as a bosonic sum, however examining the bosonic sum we found using the Beta method in section 3 eq. (\ref{3.33}) one can immediately note that the sum appearing here is of a slightly different form. More specifically, the bosonic sum, eq. (\ref{7.3}), includes an additional sum over $l=0,...,i$. This implies that, for our conjecture to be true, the sum for $A_r$ should behave as a telescopic sum in $l$ for any $i$ so that all terms apart from two will cancel. Following this intuition let us first define,
\begin{eqnarray}\label{7.5}
G_r(\lambda_{s,l},h_{s,l}, \tilde{Q}_l)=(q)^{-r}_{\infty}  \sum^{\infty}_{\{n_i\}=-\infty    \atop n \in \tilde{Q}_l+2M_{\tilde{A}_{r}}}
q^{\frac{1}{4}n^2-\frac{1}{2}n\tilde{\lambda}_{s,l}+h_{s,l}}.
\end{eqnarray}
This is motivated by the following symmetry,
\begin{eqnarray}\label{7.6}
G_r(\lambda_{s,l},h_{s,l},\tilde{Q}_l)=G_r(\lambda_{s,l}+\beta,h_{s,l}+\frac{1}{2}\beta\lambda_{s,l}+\frac{1}{4}\beta^2,\tilde{Q}_l+\beta)
\end{eqnarray}
which is a simple consequence of translating $n$ by $\beta$ where $\beta \in M_{\tilde{A}_{r}}$.
Let us now examine, on one hand, our expression for the $A_r$ diagrams,
\begin{eqnarray}\label{7.7}
A_r(\omega_i,Q)=\sum_{s=0,1} \sum_{l=0}^{i}   (-1)^sG_r(\lambda_{s,l},h_{s,l},\tilde{Q}_l).
\end{eqnarray}
While, on the other hand, our expression for the $SU(r)$ string functions, eq. (\ref{3.33}), in terms of $G_r$ is given by
\begin{eqnarray}\label{7.8}
 H^{\omega_i}_Q=\sum_{s=0,1} (-1)^sG_r(\tilde{\Lambda}_{s},d_{s},\tilde{Q}).
\end{eqnarray}
Indeed, for these two sums to be equivalent one should find that in eq. (\ref{7.7}) all $G_r$'s cancel but one for each value of $s$. Explicitly, we would like to show
\begin{eqnarray}\label{7.9}
G_r(\lambda_{1,l},h_{1,l},\tilde{Q}_l)=G_r(\lambda_{0,l+1},h_{0,l+1},\tilde{Q}_{l+1}).
\end{eqnarray}
Following the symmetry eq. (\ref{7.6}) , this equality holds if
\begin{eqnarray}\label{7.10}
\lambda_{0,l+1}-\lambda_{1,l}=\tilde{Q}_{l+1}-\tilde{Q}_{l} \mod 2M_{\tilde{A}_{r}}, \\
\nonumber h_{0,l+1}-h_{1,l}=\frac{1}{4}\lambda_{0,l+1}^2-\frac{1}{4}\lambda_{1,l}^2. \hspace{0.6in}
\end{eqnarray}
For this purpose, first consider
\begin{eqnarray}\label{7.11}
\lambda_{0,l+1}-\lambda_{1,l}=\tilde{\omega}_{i-l-1}-\tilde{\omega}_{i-l}-\tilde{\omega}_{r-1}+3\tilde{\omega}_{r}.
\end{eqnarray}
Next, using the extended cartan matrix eq. (\ref{1.1}) and our definition for $\tilde{Q}_{l}$ one can easily find,
\begin{eqnarray}\label{7.12}
\tilde{Q}_{l+1}-\tilde{Q}_l=\sum_{j=i-l}^{r-1} \tilde{\alpha}_j=\tilde{\omega}_{i-l}-\tilde{\omega}_{i-l-1}+\tilde{\omega}_{r-1}-3\tilde{\omega}_{r},
\end{eqnarray}
so that  
\begin{eqnarray}\label{7.13}
\tilde{Q}_{l+1}-\tilde{Q}_l=\lambda_{1,l}-\lambda_{0,l+1},
\end{eqnarray}
and the first condition in eq. (\ref{7.10}) is satisfied. To verify the second condition note,
\begin{eqnarray}\label{7.14}
h_{0,l+1}-h_{1,l}=Q_r-1-\frac{1}{2}l.
\end{eqnarray}
While the RHS is easily calculated by using,
\begin{eqnarray}\label{7.15}
\lambda_{0,l+1}^2-\lambda_{1,l}^2=-(\lambda_{1,l}+\lambda_{0,l+1})(\tilde{Q}_{l+1}-\tilde{Q}_{l})
\end{eqnarray}
and $\tilde{\alpha}_i \cdot \tilde{\omega}_j=\delta_{ij}$. Indeed, one finds that the second condition is also satisfied so that eq. (\ref{7.9}) holds and we arrive at the following expression for the non zero momenta diagrams,
\begin{eqnarray}\label{7.16}
A_r(\omega_i,Q)= G_r(\lambda_{0,0},h_{0,0},\tilde{Q}_0)-G_r(\lambda_{1,i},h_{1,i},\tilde{Q}_i).
\end{eqnarray}
Trivially, the first term appearing here is the same as the first term appearing in eq. (\ref{7.8}) as,
\begin{eqnarray}\label{7.17}
\lambda_{0,0}=\tilde{\Lambda}_0, \hspace{0.3in} h_{0,0}=d_0,  \hspace{0.3in}  \tilde{Q}_0=\tilde{Q}.
\end{eqnarray}
That the second term is equivalent to the one in eq. (\ref{7.8}) simply follows from the symmetry discussed above. First, consider
\begin{eqnarray}\label{7.18}
\lambda_{1,i}-\tilde{\Lambda}_1=\tilde{\omega}_i+i(\tilde{\omega}_{r-1}-3\tilde{\omega}_r).
\end{eqnarray}
To show that the first condition is satisfied, we write $\tilde{Q}_{i}-\tilde{Q}_0$ as a telescopic sum and use eq. (\ref{7.12}),
\begin{eqnarray}\label{7.19}
\tilde{Q}_{i}-\tilde{Q}_0=\sum_{l=0}^{i-1}(\tilde{Q}_{l+1}-\tilde{Q}_l)=\tilde{\omega}_i+i(\tilde{\omega}_{r-1}-3\tilde{\omega}_r).
\end{eqnarray}
To verify the second condition note that in terms of the simple roots we have,
\begin{eqnarray}\label{7.20}
\tilde{Q}_{i}-\tilde{Q}_0=\sum_{l=1}^{r-1}Min(i,l)\tilde{\alpha}_{l}.
\end{eqnarray}
Using this expression one can derive
\begin{eqnarray}\label{7.21}
\lambda_{1,i}^2-\tilde{\Lambda}_1^2=(\lambda_{1,i}+\tilde{\Lambda}_1)(\tilde{Q}_{i}-\tilde{Q}_0)=i(i-1)+2i\tilde{\Lambda}_1,
\end{eqnarray}
On the other hand, $h_{1,i}-d_1$ is calculated via the definitions eqs. (\ref{7.2} \& \ref{7.4}). To conclude we find,
\begin{eqnarray}\label{7.22}
\lambda_{1,i}-\tilde{\Lambda}_1=\tilde{Q}_{i}-\tilde{Q}_0  \\
 \nonumber h_{1,i}-d_1=\frac{1}{4}(\lambda_{1,i}^2-\tilde{\Lambda}_1^2)  
\end{eqnarray}
So that,
\begin{eqnarray}\label{7.23}
G_r(\lambda_{1,i},h_{1,i},\tilde{Q}_i)=G_r(\Lambda_1,d_1,\tilde{Q}),
\end{eqnarray}
and we arrive at the conjectured equivalence between $SU(r+1)$ string functions and $SU(r+1)$ $q$-diagrams,
\begin{equation}\label{7.24}
 \includegraphics[width=0.5\linewidth,keepaspectratio=true]{21s}
\end{equation}

\section
{\large{Discussion }}

As we have pointed out along the way the discussion above can be generalised to construct infinitely many new series of identities of multiple sums. They also provide an alternative derivation for some well known one sum identities, let us now sketch how this is done by considering a few interesting cases. The main point of the this discussion is to try to clarify the mathematical and physical interpretations of $q$-diagrams, as such, we will mainly use diagrammatic arguments and not give a full presentation of the mathematics. \\
Let us first consider the most general two node diagram, one could try to solve the diagram by first preforming the sum over the first node using the Euler identity we find,
\begin{eqnarray}\label{8.1}
A_2(b,c,Q_w,Q)=\Psi \sum_{n =Q/2 mod \mathbb{Z}} \frac{q^{\frac{3}{2}n^2-\frac{1}{2}n(b+2c)-d(b,Q)}}{(q)_{2n}} (wq^{\frac{1}{2}(1+b+Q)})_{n-\frac{Q}{2}}(-w)^{n-\frac{Q}{2}}
\end{eqnarray}
Lets consider this sum, in the sections above we have only calculated this diagram for $b$ or $c$ equal to zero, nonetheless it should be clear that one can solve either nodes in the diagram thus we can produce the result for both $c=0$ or $b=0$. The origin of this family of identities arising from this diagram can be traced, to the best of our knowledge, to Rogers. Actually, we know of at least three identities which arise for simple values of $a$, $c$ and $Q$ that appear in page 17 of \cite{RRS} which obey $b+Q=1$ \footnote{For, example $b=1$ $c=Q=0$}. Using the identities found above one can reproduce these identities and generalize this series at least for either $c=0$ or $b=0$. \\
A natural question in our context is whether one can give some CFT interpretation for the two node diagram. In particular, consider the case $c=b=Q=o$, one might be tempted to interpret the $\Psi$ appearing here as the character of the "integrated out" fermion. Actually, it is given by 
\begin{eqnarray}\label{8.2}
\Psi = (wq^{1/2})_{\infty}
\end{eqnarray}
i.e the one node which is again just a fermion. At this point the reader might conclude that, for this to make any sense, the second term in the sum above should also be associated with some CFT character. Furthermore, since to begin with we started with a fermion and the second minimal model, one expects that this CFT should be the second minimal model with $c=7/10$. Indeed, as they should, these identities arising from the decomposition of the two node $q$-diagram, precisely matches the characters of the second minimal model. Clearly, this means that we are simply decomposing the lattice minimal model theory we have constructed in section $3$ using the boson construction provided by the beta method, albeit, from the so called fermionic side of the GRR. What can be learned from this process, to answer this question consider the beta method procedure we have used in section 3. This is a familiar story, locality in the form of modular invariance highly restricts the allowed solutions, in other words the beta method is telling one how to couple the different theories in such a way that one gets a bona fide CFT.  Next say we start with an interacting bosonic CFT, can  one decompose this model as to get a well defined CFT. This is a much more difficult question,  however, clearly if this interacting theory "origin" is a product theory then such a decomposition is possible, indeed, this has allowed us to decompose or construct the bosonic characters sector by sector. To conclude, the beta method, for product theories with bosonic type characters, provides with a two way renormalisation flow in the space of theories with bosonic characters. With this in mind let us observe our results, we have found that by decomposing a node out of a $q$ diagram we were able to extract a fermionic character. This consideration seem to imply that for those theories for which the characters can be described by any connected or non connected $q$-diagram a "fermionic" renormalization flow is possible and is described by decomposing the $q$ diagram. Of course this suggestion needs to be carefully examined and a good place to start would be to try and mimic the beta method construction in a fermionic fashion using the $q$-diagram construction.



\end{document}